\documentclass[12pt]{iopart}

\usepackage{graphicx}
\usepackage{iopams}
\usepackage{physics}
\usepackage{xcolor}

\def\barray{\begin{eqnarray}}
\def\earray{\end{eqnarray}}
\def\beq{\begin{equation}}
\def\eeq{\end{equation}}

\def\id{{\bf 1}}

\begin{document}

\begin{flushright}
IFT-125-2020
\end{flushright}

\title{Integrability and scattering of the boson field theory on a lattice}

\author{Manuel Campos,  Esperanza L\'opez and Germ\'an Sierra }

\address{Instituto de F\'{\i}sica Te\'orica, Universidad Aut\'onoma de Madrid, Cantoblanco, Madrid, Spain}

%\eads{\mailto{javi.molina@upct.es},
%\mailto{german.sierra@uam.es}}

\begin{abstract}
A free boson on a lattice is the simplest field theory one can think of.
Its partition function can be easily computed in momentum space.
However,  this straightforward solution hides its integrability properties.
Here, we use the methods of exactly solvable models, that
are currently applied to  spin systems, to a  massless and massive
free boson on a 2D lattice. The Boltzmann weights of the model
are shown to satisfy the Yang-Baxter equation with a uniformization
given by  trigonometric functions in the massless case, and Jacobi
elliptic functions in the massive case. We diagonalize the row-to-row transfer matrix,
derive the conserved quantities,  and implement the quantum inverse scattering method.
Finally, we construct two  factorized scattering $S$ matrix models  for continuous degrees of freedom
using  trigonometric and elliptic functions.
These results place the free boson  model in 2D  in the same position as the rest of the models
 that are exactly solvable \`a la Yang-Baxter, offering possible applications in quantum computation.
\end{abstract}

%\pacs{03.65.-w, 02.30.Tb, 03.65.Ge, 03.65.Sq}

%\vspace{2pc}

%\noindent{\it Keywords\/}: Solvable lattice models, Correlation functions, Conformal field theory

\maketitle
%\tableofcontents

%%%%%%%%%%%%%%%%%%%%%%%%%%%%%%%%%%%%%%%%%%%%%%%%%%%%%%%%%%%%%%%%%%%%%%%

%%%%%%%%%%%%%%%%%%%%%%%%%%%%%%%%%%%%%%%%%%%%%%%%%%%%%%%%%%%%%%%%%%%%%%%
\section{Introduction}
\label{sec:intro}

Exactly Solvable Models in Statistical Mechanics and Condensed Matter Physics
have played a key role in the study of  low dimensional
many body systems \cite{B82}-\cite{M10}.
%KB93,G95,GR96,S04,DP04,M10}.
Together with field theoretical techniques,
such as Conformal Field Theory
\cite{T95}-\cite{G04},
%\cite{T95,GN98,G04},
and  numerical methods based on Tensor Networks  \cite{W92}-\cite{O19},
they have led to a precise description
of non perturbative phenomena as the
 fractionalization of the spin in antiferromagnetic spin chains and the
spin-charge separation in one dimensional metals.
Exactly Solvable Models have also appeared in the  AdS/CFT duality in the form of  spin chain
Hamiltonians  that describe the  dilation operator of the  $N=4$ super Yang-Mills theory
\cite{MZ03}-\cite{HL05}.
%\cite{MZ03,BS03,HL05}.
More recently, the algebraic Bethe ansatz has been formulated using  Tensor Networks
that allow for the application of novel  numerical techniques and possible extensions to 2D   \cite{MK12,CM15}.
The list of Exactly Solved Models is rather large: Ising, Potts,  XX, XXZ and XYZ spin chains, Hubbard,  $t-J$, etc.
They are all characterized by Hamiltonians that  commute with an infinite number of conserved quantities in involution.
These operators can be derived from the Boltzmann weights of the corresponding partition functions that
 satisfy  the Yang-Baxter equation.

The aim of this paper is to study the integrability of a free boson in the two dimensional square lattice.
This model is solvable by elementary  techniques like Fourier analysis if there is translational invariance.
However, as far as we know, its integrability has not been studied using the tools of Exactly Solvable
models like the Bethe ansatz or the Quantum Inverse Scattering method that rely on the Yang-Baxter equation.
In the models mentioned above the local degrees of freedom are discrete, e. g. spin in the Ising or
 XXZ models, fermions  in the Hubbard model, etc. In the boson model  we have to deal with continuous degrees of freedom
 given by the real values of the scalar field. Despite of this fact, we shall show that the techniques mentioned above
 can be applied directly  obtaining new knowledge  about this fundamental  model in Statistical Mechanics and Quantum Field Theory.

Another topic that we address in this paper  is the construction of  factorized scattering models  using the
Boltzmann weights of the free boson on a lattice. The former  models  describe the elastic scattering of particles,
typically solitons, in  a relativistic quantum field theory with an infinite number of conserved quantities.
The scattering of these  particles can be factorized into the product of two-particle scattering amplitudes
that, for consistency, satisfy the Yang-Baxter equation. It turns out that some  solutions of the Yang-Baxter
equation  can be used  as Boltzmann weights of a Statistical Mechanical model or,  alternatively,
 as  scattering $S$  matrices in a relativistic quantum field theory with the appropriate identifications
 of variables.   Well known
examples  of this dual application  are the 6- vertex model versus  the sine-Gordon model,  and the
Baxter's  8-vertex model  \cite{B72} versus  the Zamolodchikov's  elliptic sine-Gordon model \cite{Z79}. We shall show
that the Boltzmann weights of the boson model can be promoted to scattering $S$ matrices
with the special feature that the particles carry a continuous degree of freedom,  unlike the
more common models where it is discrete.  We construct two $S$ matrix models, one
using the trigonometric functions, and another using Jacobi  elliptic functions. Interestingly, they are similar to those proposed
by   Mussardo and Penati for the elliptic version of the sinh-Gordon model \cite{M00}.

The paper is organized as follows. In Section 2 we show that the R-matrix associated to the discretized free boson theory satisfies the Yang-Baxter equation, both for the massless and massive cases. The row-to-row transfer matrix is explicitly constructed in Section 3. From the diagonalization of the transfer matrix we recover the spectrum of the theory and obtain the expectation values of a tower of mutually commuting charges. Section 4 is devoted to the Quantum Inverse Scattering Method.
We find that the R-matrix formally coincides with the euclidean propagator of a harmonic oscillator. Using this result, we propose
an operator expression for the conserved charges. A relativistic S-matrix satisfying the axioms of factorized scattering theory is constructed in Section 5. Section 6 contains our conclusions. The paper ends with several Appendices were technical details avoided in the main body are presented.

%%%%%%%%%%%%%%%%%%%%%%%%%%%%%%%%%%%%%%%%%%%%%%%%%%%%%%%%%%%%%%%%%%%%%%%

\section{The boson field theory model}

%We consider a real variable $\phi_{ij} \in \mathbb{R}$, living on a 2D  lattice with periodic boundary conditions.
We consider a free scalar of mass $m_0$ living on a 2D  lattice with periodic boundary conditions.
%At each lattice site lives a variable $\phi_{ij} \in \mathbb{R}$.
The euclidean partition function of the model is
\beq
Z= \int \prod_{ij} d \phi_{ij} \; e^{-{1 \over 2} \; \sum_{ij}  a_x a_\tau
 \left[  \frac{ (\phi_{ij}-\phi_{i+1 j})^2}{a_x^2}  \,+\,  \frac{ (\phi_{ij}-\phi_{i j+1})^2}{a_\tau^2}  \,+\, m^2_0   \phi_{ij}^2 \right]} \; ,
\label{lZ}
\eeq
where $a_x$ and $a_\tau$  denote the lattice spacings in the spatial and euclidean time directions,  and $\phi_{ij} \in \mathbb{R}$.
The interactions described by \eqref{lZ} are pairwise between the variables  at neighbour lattice sites.
We shall reformulate this partition function as that of a vertex model in Statistical  Mechanics.
The variables  will live on the edges and the interactions take place at the vertices of a lattice
whose orientation is $45^\circ$ degrees rotated with respect to the original one
%obtained  by rotating the original one by $45^\circ$ degrees
%Setting $a_x = a_\tau=1$, we define the statistical weights on
%the tilted  lattice,
%\beq
%W({\phi_i})= e^{-{1 \over 2} \sum_{i=1}^4 \big[(\phi_i -\phi_{i+1})^2 \, + \, {m^2_B  \over 2} \phi_i^2\big] } \, ,
% W({\phi_i})= e^{-{1 \over 2} \sum_{k=0}^1 \big[
%     (\phi_{i+k,j+1} -\phi_{i+k,j})^2
%     \, + (\phi_{i+1,j+k} -\phi_{i,j+k})^2
%   \big]
%   \, - \, \sum_{k,l=0}^1 {m^2 \over 4} \phi_{i+k,j+l}^2 } \, .
%\label{W}
%\eeq
%with the fields relabelled as follows
\begin{equation}
	  \vcenter{\hbox{\includegraphics[width=0.45\textwidth]{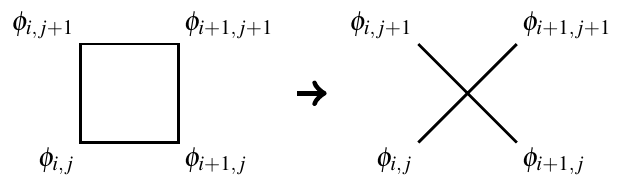}}}
		\label{figinitial}
		\end{equation}
\noindent
This model was studied in reference \cite{CS19} using the Tensor Network Renormalization
that combines renormalization group ideas with quantum information techniques.
The analysis was carried out for an isotropic lattice with $a_x = a_\tau=1$.
The continuum limit of \eqref{lZ}  is the standard partition function of a
massive boson in 2 euclidean dimensions.

\subsection{Yang--Baxter equation: massless case}

We will revisit the integrability properties of the discretized boson model using the standard techniques of Exactly Solvable Models.
In this section we shall %show that the weights associated to the free boson theory satisfy the Yang-Baxter equation,
focus on the massless case and show that it satisfies the Yang-Baxter equation. The Boltzmann weights described in \eqref{figinitial} allow to define a map
${\bf R}:\mathbb{R} \otimes \mathbb{R} \rightarrow \mathbb{R} \otimes \mathbb{R}$, known as R-matrix. R-matrices
depend on a variable that parameterizes a 1-dimensional family of models, ${\bf R}\equiv {\bf R}(c)$, and which is crucial to formulate the Yang-Baxter equation.
In order to study the free boson on a lattice from this point of view,
we need to identify a variable playing such a role.

Typically R-matrices trivialize for some value of $c$, which we will take to be $c=0$. Namely ${\bf R}(0)= {\bf I}$.
This motivates the simple choice $c={a_\tau \over a_x}$ and the definition
\begin{equation}
\label{R}
		R_{x_1 x_2}^{y_1 y_2}(c)
	= \frac{1}{2 \pi c}
		e^{ -{1 \over 2} \bqty{
			{1 \over c}   (x_1 - y_1)^2
			+ {1 \over c}  (x_2 - y_2)^2
			+ c (x_1 - x_2)^2
			+ c  (y_1 - y_2)^2
			}
			} \ .
\end{equation}
The field variables in the vertex \eqref{figinitial} have been renamed as $x_{1,2}$ and $y_{1,2}$ for simplicity. A normalization factor has been added such that
%For clarity, in the following we will use bold face letters to refer to operators and regular letter to describe their components.
%which combines the vertex \eqref{figinitial} with a  normalization factor such that
%The exponential term coincides with the vertex $W$ in \eqref{figinitial}, derived from the bosonic partition function. A normalization factor has been added such that
\beq
\lim_{c \rightarrow 0} R_{x_1 x_2}^{y_1 y_2}(c)= \delta(x_1\! -\! y_1) \delta(x_2\! -\! y_2) \ ,
%\lim_{c \rightarrow 0}  \frac{1}{2 \pi c} 		e^{ -{1 \over 2c} (  (x_1 - y_1)^2 +   (x_2 - y_2)^2 }
%{\bf I}_{x_1,x_2}^{y_1,y_2}(c) = \delta(x_1\! -\! y_1) \delta(x_2\! -\! y_2) \ ,
\label{Id}
\eeq
since $ {1\over \sqrt{2 \pi c}} e^{ - {\;\,x^2 \over 2 c}}$ approaches  a delta function as $c$ vanishes.
%For clarity,
In order to avoid confusion we will always use boldface letters to refer to operators and regular letters to describe their components.
The R-matrix components can be represented graphically as
\begin{equation}
	  \vcenter{\hbox{\includegraphics[width=0.14\textwidth]{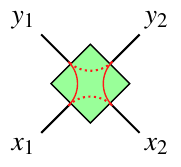}}}
	  \ ,
	  \label{tensor}
\end{equation}
with the internal lines in the green shaded vertex corresponding to the four terms in the exponent of \eqref{R}. The vertical lines represent
the terms multiplied by $c^{-1}$, and the dotted horizontal ones those multiplied by $c$.
%In the extreme anisotropic limit  $c  \rightarrow 0$ the blue lines do not contribute. Indeed, we have
In the extreme anisotropic limit \eqref{Id}, the links associated to the horizontal  lines disappear.

%and represent it by the following picture
Exactly solvable vertex models in Statistical Mechanics are those whose  Boltzmann weights
satisfy the Yang-Baxter equation (YBE).
This equation guarantees the existence of commuting row-to-row transfer matrices
whose expansion in a variable generates an infinite number of conserved quantities.
The YBE for the boson model reads
%(\textcolor{red}{brief explanation of the YB equation})
%
\beq
( {\bf R}(c_3) \otimes {\bf I} ) ({\bf I} \otimes  {\bf R}(c_2) )  ( {\bf R}(c_1) \otimes {\bf I} ) =
({\bf I} \otimes  {\bf R}(c_1) )  ( {\bf R}(c_2) \otimes {\bf I} ) ({\bf I} \otimes  {\bf R}(c_3) )  \, .
\label{YB2}
\eeq
Its  graphical representation is
\begin{equation}
	  \vcenter{\hbox{\includegraphics{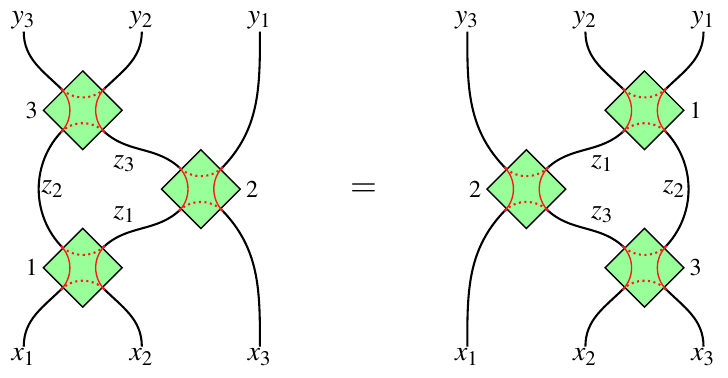}}}
  \ .
	%  \nonumber
	  \label{YBE2}
\end{equation}
It reads in components
\beq
\int  \!\!d {\vec z}  \;
R_{x_1 x_2}^{z_1 z_2}(c_1)   R_{z_2 x_3}^{z_3 y_1}(c_2)   R_{z_1 z_3}^{y_3 y_2}(c_3) = \!
\int \!\!d{\vec z}  \;
 R_{x_2 x_3}^{z_3  z_1}(c_3)   R_{x_1 z_3}^{y_3 z_2}(c_2)   R_{z_2 z_1}^{y_2 y_1}(c_1)  \ ,
\label{YB1}
\eeq
where ${\vec z}=(z_1,z_2,z_3)$ and each integration  runs over $\mathbb{R}$.
The integration  replaces the  sum over a finite set of  variables
in the standard spin models.

We shall next determine the conditions that the parameters $c_{i=1,2,3}$ must verify in order to fulfill \eqref{YB1}.
The weights \eqref{R}  satisfy
\beq
R_{x_1 x_2}^{y_1 y_2}(c) = R_{y_2 y_1}^{x_2 x_1}(c)  \, ,
\label{reflex}
\eeq
which is equivalent to the invariance of the vertex \eqref{tensor} under a $180^\circ$ rotation.
This discrete symmetry implies that the {\it rhs} of the YBE  equals the {\it lhs} with the roles of $x$ and $y$ variables exchanged.
Therefore it is enough to require that the {\it lhs} of \eqref{YB1} defines a matrix invariant under $x_i \leftrightarrow y_i$.
%
%\beq
%\int d{\vec z}  \,
%R_{x_1,x_2}^{z_1,z_2}(c_1)  \, R_{z_2,x_3}^{z_3,y_1}(c_2)  \, R_{z_1,z_3}^{y_3,y_2}(c_3)  \ ,
%\label{YB11}
%\eeq
%
%is invariant under the exchange of $x_i$ and $y_i$ variables.
%Since the Boltzmann weights are gaussian, this expression can be rewritten as
The {\it lhs} of \eqref{YB1} can be rewritten as
\beq
\int d{\vec z}  \,
e^{ - ( \vec{ x},\vec{ y} , \vec{ z} )    M ( \vec{ x},\vec{ y} , \vec{ z} )^T }  \ , \hspace{8mm} M = \left( \begin{array}{ccc}
M_{xx} & M_{xy} & M_{xz} \\
M_{yx}  & M_{yy} & M_{yz} \\
M_{zx}  & M_{zy} & M_{zz} \\
\end{array}
\right) \, ,
\label{YB3}
\eeq
with ${\vec x}=(x_1,x_2,x_3)$ and ${\vec y}=(y_1,y_2,y_3)$, and $M$ a symmetric matrix
whose entries are blocks of size $3 \times  3$.
The explicit expression of $M$ is given in Appendix A. The $z$-integrations can be easily performed, obtaining
\beq
e^{ - ( \vec{ x},\vec{ y})    N ( \vec{ x},\vec{ y} )^T } \ ,  \hspace{.6cm}
\label{YB33}
%\eeq
%where
%
%\begin{equation}
%\label{condition}
N=
		\begin{pmatrix}
			M_{x x} & M_{x y} \\
			M_{y x} & M_{y y}
		\end{pmatrix}
		-
		\begin{pmatrix}
			M_{x z} \\
			M_{y z}
		\end{pmatrix}
		\! M_{z z}^{-1} \!
		\begin{pmatrix}
			M_{z x} & M_{z y}
		\end{pmatrix}
  \ .
\end{equation}
The YBE translates into requiring $N_{xx}=N_{yy}$ and $N_{xy}=N_{yx}$. In Appendix A we show that
these conditions hold if and only if
\beq
c_2 =  \frac{c_1 + c_3}{1- c_1 c_3}  \ .
\label{c123}
\eeq

The relation between the parameters $c_i$ is uniformized by the function  $c(u) = \tan(u)$, becoming equivalent to
\begin{equation}
\label{cu}
	c_1 = c(u)
	\ , \quad
	c_2  = c(u+v)
	\ , \quad
	c_3  = c(v)
	\ .
\end{equation}
%with $u$ and $v$ unconstrained (see below).
%Let us  assign a pair of  variables
%(\textcolor{red}{explain})
%$u_1$ and $u_2$ to the matrix \eqref{R}
%and write it as ${\bf R}(u_{12})  \equiv {\bf R}(c(u_1\! -\! u_2))$. In terms of these variables
%the YBE equation  \eqref{YB2}  reads
%
%\beq
%( {\bf R}(u_{23}) \otimes {\bf I} ) ({\bf I} \otimes  {\bf R}(u_{13}) )  ( {\bf R}(u_{12}) \otimes {\bf I} ) =
%({\bf I} \otimes  {\bf R}(u_{12}) )  ( {\bf R}(u_{13}) \otimes {\bf I} ) ({\bf I} \otimes  {\bf R}(u_{23}) )  \, .
%( {\bf R}(u_2- u_3) \otimes {\bf I} ) ({\bf I} \otimes  {\bf R}(u_1 - u_3) )  ( {\bf R}(u_1- u_2) \otimes {\bf I} ) =
%({\bf I} \otimes  {\bf R}(u_1- u_2) )  ( {\bf R}(u_1- u_3) \otimes {\bf I} ) ({\bf I} \otimes  {\bf R}(u_2 - u_3) )  \, .
%\label{YB4}
%\eeq
%This  is the familiar form of the YBE for $R$ matrices that have  the difference property,
%in the rapidity,
%that is ${\bf R}(u_1 ,  u_2)  = {\bf R}(u_1\! -\! u_2)$. Well known examples are
%the  $R$ matrices of the 6 and 8 vertex models in Statistical  Mechanics \cite{B82}.
The physical interpretation of $R(u)$ as the Boltzmann weights of the massless boson requires $c$ to be positive,
which is crucial for the regularity of the large field limit and the convergence of the integration \eqref{YB3}  over the internal $z's$ variables.
Therefore, the set where $u$ can take values has to be restricted. We chose it to be the interval $\left[ 0, \frac{\pi}{2} \right]$, such that
in the limit of small $c$ we have $c\approx u$.
%the uniformization $c=\tan u$ leads to chose $u$ in the interval $\left( 0, \frac{\pi}{2} \right)$ (\textcolor{red}{relation with S-matrix}).
%and the physical interpretation of $R(u)$ as the Boltzmann weights of the massless boson, implies that
%$u$ must be chosen in the interval $\left( 0, \frac{\pi}{2} \right)$ (\textcolor{red}{relation with S-matrix}).
%When considering the YBE \eqref{YB4}, notice that naturally $u_1 > u_2 > u_3$.
This sort of restriction does not arise in other models like the 6 vertex where the parameter
$u$ can take any complex value.

%We finally notice that the isotropic Boltzmann weights \eqref{W}, with $m=0$,
%are recovered for the  spectral parameter
%
%\beq
%u_B = \frac{\pi}{4} \rightarrow c(u_B) = 1  \ .
%\label{iso}
%\eeq

%%%%%%%%%%%%%%%%%%%%%%%%%%%%%%%%%%%%%%%%%%%%%%%%%%%%%%%%%%%%%%%%%%%%%%%
\subsection{Yang--Baxter equation:  massive case}

We introduce the following modification of the $R$-matrix
%The mass term in the Boltzmann weights   \eqref{W} corresponds to the following modification of the $R$-tensor \eqref{R}
\begin{equation}
\label{Rm}
%\tilde{R}_{x_1,x_2}^{y_1,y_2}(c,\nu)
%	=
%		R_{x_1,x_2}^{y_1,y_2}(c)
%		e^{ - {\nu \over 4} ( x_1^2 + x_2^2 + y_1^2 + y_2^2 )	}
R_{x_1 x_2}^{y_1 y_2}(c,{\widetilde m})
	=
		R_{x_1 x_2}^{y_1 y_2}(c)
		e^{ - {c \over 4} {\widetilde m}^2 ( x_1^2 + x_2^2 + y_1^2 + y_2^2 )	}
	\ ,
\end{equation}
corresponding to the massive deformation of the free boson model.
The parameter ${\widetilde m}$ is the boson mass measured in lattice units
\begin{equation}
{\widetilde m}=  a_x m_0 \ .
\label{mass0}
\end{equation}
The simple assumption of keeping $\widetilde m$ constant while the lattice anisotropy $c$ varies, fails
to satisfy the YBE. Therefore we will allow it to be a general function of $c$,
and determine it by imposing
\barray
&&\int  d{\vec z}  \,
R_{x_1 x_2}^{z_1 z_2}(c_1,{\widetilde m}_1)  \, R_{z_2 x_3}^{z_3 y_1}(c_2,{\widetilde m}_2)  \, R_{z_1 z_3}^{y_3 y_2}(c_3,{\widetilde m}_3)
\label{YBm1} \\[-1mm]
  &&\hspace{1.9cm}=
\int d {\vec z}  \,
R_{x_2 x_3}^{z_3  z_1}(c_3,{\widetilde m}_3)  \, R_{x_1 z_3}^{y_3 z_2}(c_2,{\widetilde m}_2)  \, R_{z_2 z_1}^{y_2 y_1}(c_1,{\widetilde m}_1)  \ .
\nonumber
\earray

The R-matrix \eqref{Rm} is invariant under the $180^\circ$ rotation \eqref{reflex}, implying that
the treatment of the previous section also applies to the massive deformation. Following the same steps (see Appendix A), we can show
 that the YBE holds if and only if

\vspace{-3mm}
\barray
\label{l1}
&& { c_1{\widetilde m}^2_1-c_2 {\widetilde m}^2_2 \over c_1 {\widetilde m}^2_1 -c_3{\widetilde m}^2_3} =   {c_3 -c_1 c_2 c_3 \over c_3- c_1} \; , \hspace{1cm}    {c_3 {\widetilde m}^2_3-c_2{\widetilde m}^2_2 \over c_3 {\widetilde m}^2_3 -c_1 {\widetilde m}^2_1} =   {c_1 - c_1 c_2 c_3 \over c_1- c_3 }  \; ,  \\[3mm]
\label{l3}
&& \hspace{1.2cm} {\widetilde m}^2_2 =  {(1- c_1 c_3)^2 \over c_1  c_3} \left[ 1 - {1 \over c_2^2}\left( {c_1+c_3 \over 1- c_1 c_3} \right)^{\!\! 2} \, \right] \ .
\earray

\noindent
The requirement that each ${\widetilde m}_i$ be independent of the variables $c_{j \neq i}$ is far from obvious in view of \eqref{l3}.
%As an additional physical requirement, each mass should depend only on its associated spectral parameter, $\nu_i=\nu_i(c_i)$. In view of \eqref{l3}, it is far from obvious that this condition can be met.
There is however a two parameter family of solutions which generalizes those of the massless case by promoting trigonometric to elliptic functions.
Let us define
\begin{equation}
\label{cumm}
	c(u,\mu) = \sqrt{\mu_1}\, {{\rm s n}(u,\mu) \over {\rm c n}(u,\mu)   {\rm dn}(u, \mu)}
	\ , \hspace{6mm}	{\widetilde m}(u, \mu)  = \sqrt{4 \mu \over \mu_1} \,{\rm cn}(u,\mu) \ ,
\end{equation}
%\begin{equation}
%\label{cum}
%	c(u,\mu) = {{\rm s n}(u,\mu) {\rm d n}(u,\mu)  \over {\rm c n}(u,\mu)}
%	\ ,  \quad  m(u, \mu)^2 =  - {4 \mu  \,  {\rm c n}(u,\mu)  {\rm s n} (u,\mu)   \over {\rm d n}(u,\mu)}
%\end{equation}
%
where $\mu_1 = 1- \mu$ and
${\rm sn}(u,\mu), {\rm cn}(u,\mu), {\rm dn}(u,\mu)$ are Jacobi elliptic functions of argument $u$ and parameter   $\mu$ \cite{AS}.
Equations \eqref{l1}-\eqref{l3} are satisfied by
\begin{equation}
	c_1 = c(u,\mu)
	\ , \quad
	c_2 = c(u+v,\mu)
	\ , \quad
	c_3  = c(v,\mu)
	\ ,
\end{equation}
%into \eqref{lam}  we obtain  the dependence of the masses on $u$ and $k$
\begin{equation}
	{\widetilde m}_1  = {\widetilde m}(u,\mu)
	\ , \quad
	{\widetilde m}_2   = {\widetilde m}(u+v,\mu)
	\ , \quad
	{\widetilde m}_3  = {\widetilde m}(v,\mu)
	\ .
\end{equation}
%The difference property in the rapidities $u$, $v$ also holds in the massive case.
The massless uniformization is recovered at vanishing elliptic parameter $\mu$.
As in that case, regularity of the R-matrix in the large field limit forces $u$ to
take values on a restricted interval. Asking again for a linear relation between $u$ and $c$ when $c$ is small, we choose  $u\in[0,K(\mu)]$ with
\beq
\label{K}
K(\mu) = \int_0^{\pi/2}  \frac{ d \theta}{ \sqrt{ 1 -  \mu  \sin^2 \theta} }  \ ,
\eeq
the complete elliptic integral of the first kind.
Since $K(0)={\pi \over 2}$, we conclude that the YBE is compatible with a smooth deformation away from the massless case.

\begin{figure}[t!]
\vspace{.2cm}
\begin{center}
\includegraphics[width=0.4\textwidth]{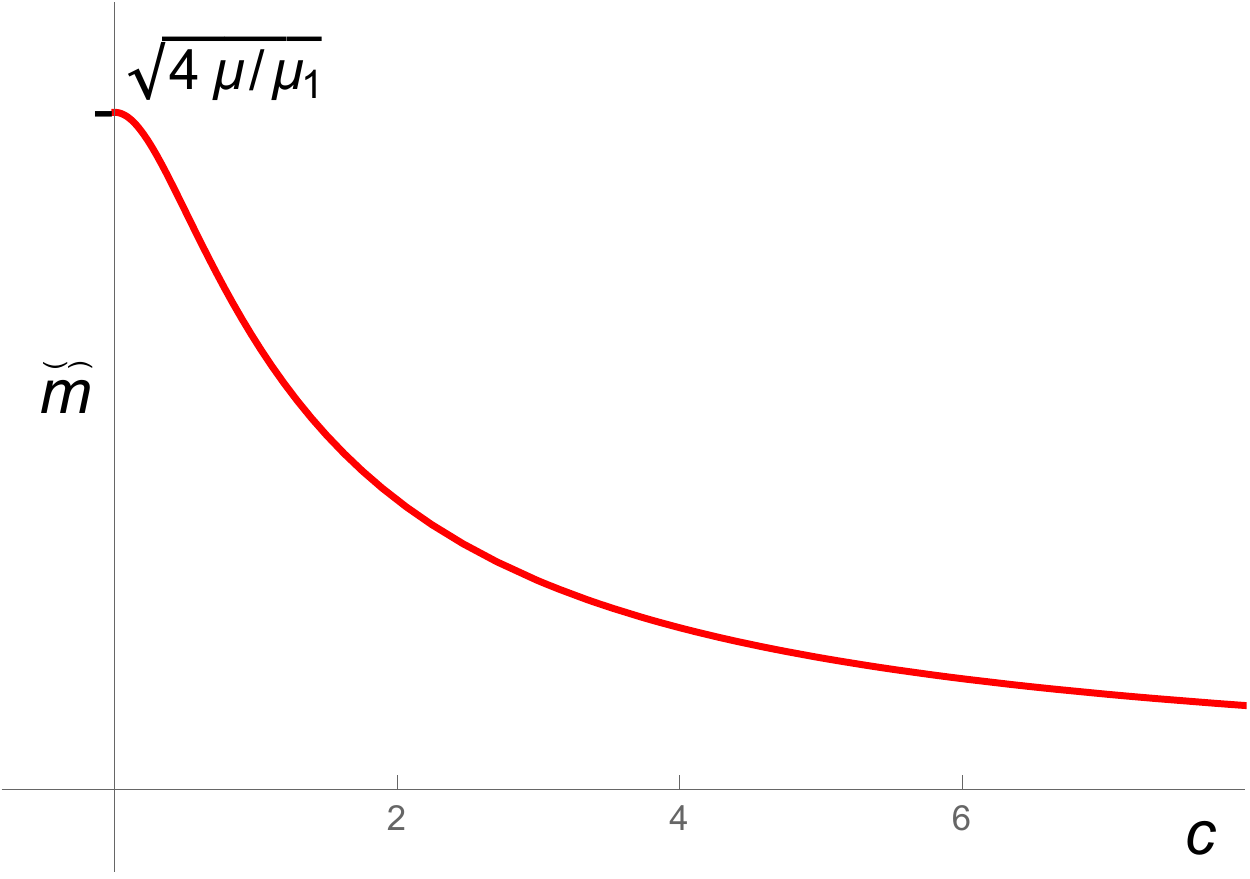} \qquad
\includegraphics[width=0.4\textwidth]{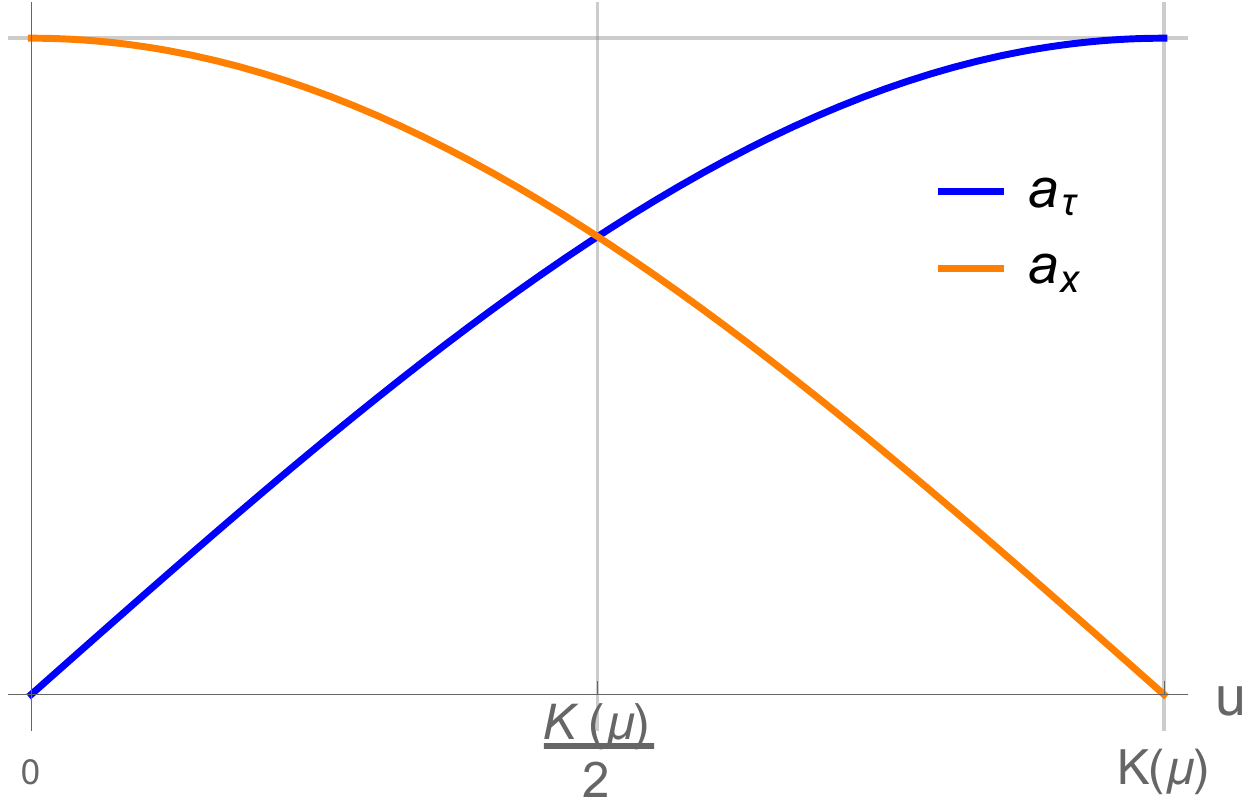}
\caption{
Left: Family of lattice models described by \eqref{cumm}. Right: Dependence of the lattice steps $a_\tau$ and $a_x$ on the uniformization parameter, under the assumption that the boson mass is constant along the family.
}
\label{mass}
\end{center}
\end{figure}

Fig.\ref{mass} shows the parameter space of the discretized bosonic theory together with the 1-dimensional family of models selected by the YBE. Attributing the
variation of $\widetilde m$ entirely to the lattice step $a_x$ included in its definition \eqref{mass0}, the following consistent picture is obtained. The
boson mass keeps constant along the family of models \eqref{cumm}. The
dependence of the spatial and temporal lattice steps on the uniformization parameter is
then easily obtained, and have been plotted in Fig.\ref{mass}. Although $c$ diverges as $u \to K(\mu)$, both $a_x$ and $a_\tau$ remain finite. Moreover for $u$ larger than ${K(\mu) \over 2}$, the point at
which the lattice becomes isotropic, the roles of the spatial and euclidean time directions get exchanged.
We identify the boson mass as, $m_0 \propto \sqrt{4 \mu \over \mu_1}$.
%We associate the square root factor in the expression \eqref{cumm} of $\widetilde m$ as a function of $u$ with the boson mass, $m \propto \sqrt{4 \mu \over \mu_1}$.
Positive $m_0$ requires $\mu \in (0, 1)$, with $\mu \to 1$ corresponding to the infinite mass limit.

%
%\begin{figure}[t!]
%\vspace{.2cm}
%\begin{center}
%\includegraphics[width=0.5\textwidth]{mass}
%\caption{
%Plot of the masses $(a_x m_0)^2$ (red curve)  and $(a_x m_B)^2$ (blue curve)  given in  equations \eqref{axm} and  \eqref{m2}
%respectively for   $\mu \in (0, 1)$.
%  }
%\end{center}
%\label{mass}
%\end{figure}

%%%%%%%%%%%%%%%%%%%%%%%%%%%%%%%%%%%%%%%%%%%%%%%%%%%%%%%%%%%%%%%%%%%%%%%
\section{The row-to-row transfer matrix}\label{coordinateTM}
%%%%%%%%%%%%%%%%%%%%%%%%%%%%%%%%%%%%%%%%%%%%%%%%%%%%%%%%%%%%%%%%%%%%%%%

%\subsection{Exact expression}

%\textcolor{red}{bla bla tranfer matrix}
The partition function $Z$  of a vertex model on a $L \times N$  lattice can be written
as $Z= {\rm tr} \;  T^N$, where $T$ is the product of Boltzmann weights along a row with their horizontal variables identified and summed over,  including the first and the last ones. This defines the so called
row-to-row transfer matrix that for  the massive  boson model is  the map   ${\bf T}(u): {\cal H} \to {\cal H}$
%\mathbb{R}^{\otimes L} \to \mathbb{R}^{\otimes L}$
depicted as

\begin{equation}
	  {\bf T}(u) =
		\vcenter{\hbox{\includegraphics{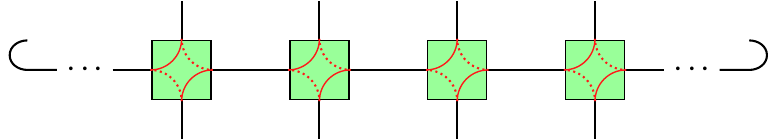}}} \ .
	  \label{Tpicture}
\end{equation}
where ${\cal H}= \mathbb{R}^{\otimes L}$
is the lattice Hilbert space and the same variable  $u$ characterizes every vertex.

The matrix elements of the transfer matrix read
\begin{equation}
\label{T}
		\langle {\vec y} |\, {\bf T}(u) | {\vec x} \rangle \equiv T(\vec x, \vec y;u)
	=
		{1\over (2 \pi c)^L}  \int   \dd {\vec z}   \,
			e^{ - {1 \over 2}( \vec{ x},\vec{ y} , \vec{ z} )  \,  M \, ( \vec{ x},\vec{ y} , \vec{ z} )^T }  \ ,
\end{equation}
where $\vec x$ are the bottom variables, $\vec y$ the top ones and $\vec z$ the variables  associated
to the horizontal line. %,  that are integrated over.
$M$ is a  symmetric matrix analogous to \eqref{YB3} but whose blocks are
of dimension $L \times L$ (see Appendix B).
%The diagonal blocks are
%\begin{equation}
%		M_{x x} = M_{y y} = {1 \over 2 } M_{zz} = a \, \id
%	\ , \quad   a= {1 \over c} + c + {\nu \over 2} \ ,
%\end{equation}
%and the off-diagonal blocks
%\begin{equation}
%		M_{xy}=0 \ , \quad
%		M_{x z} = - {1 \over c} \big( \id - c^2  S \big)
%	\ , \quad
%	M_{y z} =  - {1 \over c} \big( S - c^2 \id \big)
%			\ .
%\end{equation}
%The shift matrix is defined as $S_{ij}=\delta_{i,j+1}$  with $L+1 \equiv 1$.
The integration on the $z$-variables can be easily done, leading to
an explicit expression for the transfer matrix
\begin{equation}
\label{T2}
		 T(\vec x, \vec y;u)
	=
		{ 1 \over (4 \pi a c^2)^{L\over 2}}    \,
		e^{ - {1 \over 2}( \vec{ x},\vec{ y} )  \,  N \, ( \vec{ x},\vec{ y}  )^T }  \ ,
\end{equation}
where we have introduced the convenient combination
$a= {1 \over c} + c + {c \over 2}{\widetilde m}^2$,
and $N$ is again a symmetric matrix
\barray
N_{xx}=N_{yy}=\left( a- {1 + c^4  \over 2 a c^2}\right) \id - {1\over 2a} \big(S + S^T \big)\equiv N_1 \label{T21} \ , \\
N_{xy}=-{1 \over a}  \, \id - {1 \over 2ac^2} \big( S + c^4 S^T \big) \equiv N_2 \label{T22} \ ,
\earray
with the shift matrix $S_{ij}=\delta_{i,j+1}$  and $L+1 \equiv 1$. We have defined the matrices $N_1$ and $N_2$ for latter convenience.

%where $N={1 \over 2 a} {\widetilde N}$ and
%\begin{equation}
%	\label{M1}
%		{\widetilde N}_{x x}
%	=
%		{\widetilde N}_{y y}
%	=
%		(2a^2\! -\! c^2 \!- \!c^{-2}) \id - (S + S^T)
%	\ , \quad
%	{\widetilde N}_{x y}
%	=
%		- 2 \, \id - ( c^{-2} S + c^2 S^T ) \ .
%\end{equation}

%\subsection{Conserved charges}

In order to better understand how the tower of conserved charges emerges from the expansion of the transfer matrix, it is convenient
to rewrite its components as a product of two factors, $T=T_p \,T_q$, with
\barray
%\hspace{-.1cm} T_0 & = & g^L  \ ,\\
T_p (\vec x, \vec y;u) &\! \!=\!\!&{ 1 \over (4 \pi a c^2)^{L\over 2}} e^{ -{1 \over 4 a c^2} \sum_i (x_i-y_{i+1})^2} \ , \label{Tpp}  \\[1mm]
 T_q (\vec x, \vec y;u) & \!\!=\! \!& e^{-{1 \over 4 a} \sum_i \left[ {8\mu \over \mu_1} (x_i^2+ y_i^2) +2 (x_i-y_i)^2+(x_i-x_{i+1})^2 +(y_i-y_{i+1})^2  +c^2 (x_i-y_{i-1})^2 \right]}  \label{Tqq} \; \;
\earray
%each with a different behaviour in the limit $\theta \rightarrow 0$. The global factor $T_0$ is non-analytic at zero rapidity but it has a well defined value, $T_0(0)=1$. It has no influence
%in the derivation of the conserved charges and we will disregard it in the following.
The function $T_p$ tends to a delta distribution as $u \rightarrow 0$.
At small $u$, $T_p$ can be expanded in terms of derivatives of a delta function
(see Appendix B)
%\begin{equation}
%T_1(\vec x, \vec y) =  \delta(\vec x- S \vec y) \, e^{a c^2 \sum_i D_i^2}  \ , \quad D_i= {\partial\over \partial x_i}-{\partial \over \partial y_{i+1}}
%\end{equation}
\begin{equation}
T_p(\vec x, \vec y;u) = \prod_{i=1}^L \, \left( \, \sum_{k=0}^\infty { (a c^2)^k \over k! } \delta^{(2k)} (x_i \!-\! y_{i+1}) \right) \ ,
%T_p(\vec x, \vec y;u) = \sum_{k=0}^\infty { (a c^2)^k \over k! } \delta^{2k)} ({\vec x} \!-\! S {\vec y}) \ .
\label{cute}
\end{equation}
%Using this and the analiticity of $\bar T$, we can rewrite the transfer matrix as
%\begin{equation}
%T (\vec x, \vec y) = g^L \,  \delta(\vec x- S \vec y) \, e^{a c^2 \sum_i D_i^2} \; T_2 (\vec x, \vec y) \ , \quad D_i= {\partial\over \partial x_i}-{\partial \over \partial y_{i\!+\!1}}
%\end{equation}
%This can be further recast in the concise form
%\begin{equation}
%\label{cute}
%T_p  (\vec x, \vec y;u) =  \delta(\vec x- S \vec y) \, e^{{a c^2 \over 4} \sum_i (\partial_{x_i}\!-\,\partial_{y_{i+\!1}})^2 }  \ ,
%\end{equation}
%
which as $T_q$, has a well defined expansion around $u=0$.
At leading order only the $k=0$ term in \eqref{cute} contributes to the transfer matrix. Hence it reduces to a cyclic permutation
\begin{equation}
%{\bf T}(0)
e^{i \bf P} |x_1,  x_2 \dots, x_L\rangle  = |x_L, x_1,  \dots, x_{L-1} \rangle \ .
\label{Tt0}
\eeq
where $a_x^{-1} {\bf P}$ is the lattice momentum, the first conserved charge derived from the transfer matrix expansion.
A graphical derivation of the identification ${\bf T}(0)=e^{i \bf P}$ can be obtained from \eqref{Tpicture}, since at $u$ or equivalently $c=0$, the dotted links inside the vertices vanish.
%This is also clear from the graphical representation \eqref{Tpicture}, since at $u$ or equivalently $c=0$, the blue links inside the vertices vanish.
%For simplicity, we will use $c$ instead of the rapidity as expansion parameter. $c$ is analytic and $\propto \theta$ at small rapidity, therefore this choice just amounts to a linear redefinition of the conserved charges. We define
%This operator can be written as $T_0 = e^{ i \hat{P}}$, where $\hat{P}$ is  the lattice momentum operator.

At higher orders $T_q$ contributes with powers of the bosonic field, and its discretized spatial derivatives to the corresponding conserved charge, while $T_p$ provides powers of the canonical momentum.
The canonical commutations of the bosonic field $\bf x$ and its conjugate momentum $\boldsymbol \pi$ in the continuum, become in the discretized model
\begin{equation}
[ {\bf x}(z) , {\boldsymbol \pi}(z')]= i \delta(z\!-\!z')\;\;  \longrightarrow \;\; [ {\bf x}_i , {\boldsymbol \pi}_j]= i a_x^{-1} \delta_{ij} \ .
\end{equation}
Therefore we can represent ${\boldsymbol \pi}_i= -i a_x^{-1} \partial_{x_i}$, where
\begin{equation}
\langle {\vec y} |\, \partial_{x_i} | {\vec x} \rangle =  \delta' (x_i \!-\!y_i) \prod_{j \neq i} \delta(x_j\!-\!y_j) \ .
\label{cpi}
\end{equation}
Using this, it is immediate to obtain the next to leading contribution to the transfer matrix
%it is immediate to obtain the Hamiltonian from the next to leading contribution in the transfer matrix expansion
\begin{equation}
%{\bf T}(u) |_{{\cal O}(u)} =
-2 \sqrt{\mu_1} u \; e^{i {\bf P}} \, {\bf H} \ , \hspace{.7cm} {\bf H}= {1 \over 2}  \sum_{i=1}^L \left( a_x^2  {\boldsymbol \pi}_i^2 + ({\bf x}_i \!-\! {\bf x}_{i+1} )^2
+ {\widetilde m}_0^2 \,
{\bf x}_i^2 \right) \ ,
%{\bf T}(u) |_{{\cal O}(u)} = - 2 \sqrt{\mu_1} u \; e^{i \bf P} \, \sum_{i=1}^L {1 \over 2} \left( {\boldsymbol \pi}_i^2 + {({\bf x}_i \!-\! {\bf x}_{i+1} )^2 \over a_x^2} + m_0^2 \,{\bf x}_i^2 \right) \ ,
\label{linearT}
\end{equation}
where $a_x^{-2}{\bf H}$ is the discretized free boson Hamiltonian and ${\widetilde m_0}^2\!=\! {4 \mu \over \mu_1}$, in agreement with \eqref{cumm}. If we interpret $a_x$ as a u-dependent parameter, as done in Fig.\ref{mass}, we should select here its value at vanishing $u$.
Obtaining the higher conserved charges along these lines is possible but cumbersome.
Below we will follow an alternative strategy.

%At first order we have
%\begin{equation}
%\label{Tt1}
%\langle {\vec y} | T_1 | {\vec x} \rangle =  -\delta^{''} \!({\vec x}- {\vec y}) + \delta ({\vec x}- {\vec y})\sum_{i=1}^L \left[ (x_{i+1}-x_i)^2 +m^2 x_i^2 \right] \ ,
%\end{equation}
%which implies that $T_1=2H$, the Hamiltonian of the free bosonic theory.

%\begin{equation}
%T_1(\vec x, \vec y) =  \delta(\vec x- S \vec y) \prod_{I=1}^L\left[ \sum_{k=0}^\infty {(2k-1)!!  \over (2k)! 2^k} (4a c^2)^k  D^k \right] \ , \quad D= {\partial\over \partial x_i}-{\partial \over \partial y_{i+1}}
%\end{equation}

%$\widetilde M$ has the following blocks
%\begin{equation}
%	\label{M1}
%		\widetilde M_{x x}
%	=
%		\widetilde M_{y y}
%	=
%		\pqty{a - {c^2 + c^{-2} \over 2a}} \id - {1 \over 2 a} (S + S^T)
%	\ ,
%\end{equation}
%\begin{equation}
%	\label{M2}
%		\widetilde M_{x y}
%	=
%		- a^{-1} \id - {1 \over 2 a} ( c^{-2} S + c^2 S^T )
%	\ .
%\end{equation}

%%%%%%%%%%%%%%%%%%%%%%%%%%%%%%%%%%%%%%%%%%%%%%%%%%%%%%%%%%%%%%%%%%%%%%%
\subsection{Spectrum}

In this section  we diagonalize the transfer matrix of the free boson lattice theory
in a way that is  reminiscent  to the coordinate Bethe ansatz for spin systems.
Since the Hamiltonian commutes with the transfer matrix, the eigenstates of the former are also eigenstates of the latter.
%In this section we will recover the spectrum of the free boson lattice theory by diagonalizing the transfer matrix.
%Let us compute the eigenstates of the transfer matrix
%\begin{equation}
 %T(c) \ket{ \Psi } = \Lambda(c) \ket{ \Psi }  \ .
%\end{equation}
A natural ansatz for the eigenstates is
\begin{equation}
\label{eigen1}
		\ket{ \Psi }
	=	\int \dd {\vec x} \,
			f_n ({\vec x}) \,e^{ - {1 \over 2} {\vec x} \, K \, {\vec x}^T } \,| {\vec x} \rangle  \ ,
\end{equation}
where $K$ a symmetric matrix and $f_n$ a polynomial of degree $n$ in $x_i$. %, both independent of $c$.
Straightforward manipulations, detailed in Appendix B, show that the eigenstate condition ${\bf T}(u) \ket{ \Psi } = \Lambda \ket{ \Psi }$ implies
%The eigenstate condition ${\bf T}(u) \ket{ \Psi } = \Lambda \ket{ \Psi }$ reads in components
%\begin{equation}
%\label{eigencond}
%{1 \over (4 \pi a c^2)^{L \over 2} }
%\int \dd {\vec x} \,  e^{ - {1 \over 2}( \vec{ x},\vec{ y} )  \,  N \, ( \vec{ x},\vec{ y}  )^T } f_n({\vec x})  \, e^{ - {1 \over 2} {\vec x}\, K\, {\vec x}^T } = {\bar \Lambda} \,f_n({\vec y}) \, e^{ - {1 \over 2} {\vec y}\, K \,{\vec y}^T } \ ,
%\end{equation}
%with ${\bar \Lambda}=(4 \pi a c^2)^{L \over 2} \Lambda$ and the matrix $N$ given in \eqref{T2}-\eqref{T22}. Recall that since $u$ does not appear in the %Hamiltonian, both $K$ and $f_n$ should be independent of it.
%Equation \eqref{eigencond} implies
%\begin{equation}
%	N_1 -  N_2^T ( N_1 + K)^{-1} N_2	= K	\ ,
%\end{equation}
\begin{equation}
\label{hermite}
{1 \over (4 \pi a c^2)^{L \over 2} }    \int \dd {\vec x} \,
		e^{ - {1 \over 2} \vec{ x}  \,  (N_1+K) \, \vec{ x}^T } f_n\big({\vec x}-{\vec y} \,N_2^T (N_1 + K)^{-1} \big) =\Lambda \, f_n({\vec y}) \ ,
\end{equation}
together with
\begin{equation}
\label{eigenmatrix}
	N_1 -  N_2^T ( N_1 + K)^{-1} N_2
	= K	\ .
\end{equation}
The matrices $N_{1,2}$ have been defined in \eqref{T21}-\eqref{T22}. Recall that, although they depend in $u$, $K$,
the function $f_n$ should be independent of the uniformization parameter.

The matrices $N_1$ and $N_2$ commute because they are linear combinations of the identity, the shift matrix and its transpose. Assuming that $K$ also commutes with them, and using that the $a$ can be rewritten as
\begin{equation}
a= \sqrt{ \Big( c+ {1 \over c} \Big)^2 + {\widetilde m}_0^2} \ ,
\label{ac}
\end{equation}
we obtain
\begin{equation}
\label{Q}
K^2 = N_1^2 - N_2^T N_2 =\left( {\widetilde m}_0^2 +2 \right) {\bf 1} -S-S^T \ ,
\end{equation}
This expression is indeed consistent with the previous assumption and satisfies the required independence of $u$. The eigenvalues of the shift matrix $S$ are roots of unity of order $L$. Hence the eigenvalues of $K$ are
%The eigenvalues of $Q^2$ are $q_l^2=4 \omega^2_l$ with
\begin{equation}
\label{omegak}
%{\pi k \over L}
\omega_k= \sqrt{{\widetilde m}_0^2 +4\sin^2 {p_k \over 2}}  \ ,  \hspace{1cm} k=0,...,L-1   \ .
\end{equation}
with $p_k={2 \pi k \over L}$.
%Notice that, although $Q^2$ only has entries on the diagonal and one step above or below it, its square root is a non-local matrix.
%The state \eqref{eigen} for $n=0$ coincides with the ground state of a collection of harmonic oscillators of frequency $\omega_k$.
%It is also the groud state of the lattice model. since the energy of a scalar on a lattice of $L$ sites with momentum $p_k= {2 \pi k \over a_x L}$ and mass
%$m_0$ is $a_x^{-1} \omega_k$.
The state \eqref{eigen1} with $f_0$ a constant is the ground state of the lattice model. Indeed, the energy of a bosonic eigenmode with
momentum $ a_x^{-1} p_k$ and mass $m_0$ is $a_x^{-1} \omega_k$.  Notice that, although $K^2$ only has entries on the diagonal and
one step above or below it, its square root is a non-local matrix.
The eigenvalue of the transfer matrix on the ground state is
\begin{equation}
\Lambda_0={ 1 \over (2 a c^2)^{L \over 2}\sqrt{\det (N_1+K)}}={ 1 \over c^L \prod_k (a+ \omega_k )} \ .
\label{L0}
\end{equation}
%where we have used that the eigenvalues of $N_1+Q$ have the following simple expression
%\begin{equation}
%{1 \over 2a} \left( a+ \omega_k  \right)^2   \ .
%\end{equation}

Excited states are obtained when the function $f_n({\vec x})$ in \eqref{eigen} is non-trivial. The obvious choice for this function are the Hermite polynomials
\begin{equation}
\label{fnhermite}
f_n({\vec x}\, ;{\vec v}) = \rho \, e^{  \, {\vec x} \, K \, {\vec x}^T } \left( \! {\vec v}.{\partial \over \partial {\vec x}}  \right)^{\!\!n} e^{ -  {\vec x} \, K \, {\vec x}^T }  \ ,
\end{equation}
with $\rho$ a normalization constant and ${\vec v}$ a vector to be determined. Substituting the above ansatz into the eigenstate condition \eqref{hermite}, we obtain (see Appendix B)
\begin{equation}
	\label{eigenvaluehermite}
\Lambda_0\,  f_n\! \left({\vec y}\, ; - {\vec v}  \, (N_1+K)^{-1} N_2\right) = \Lambda \, f_n({\vec y}\, ; {\vec v} ) \ .
\end{equation}
%\begin{equation}
%e^{{\vec y} \, K \, {\vec y}^T}  \int \dd {\vec x} \,
%		e^{ - {1 \over 2} \vec{ x}  \,  (N_1-K) \, \vec{ x}^T } \left( \! {\vec v} \cdot {\partial \over \partial {\vec x}}  \right)^{\!\!n}
%		e^{ -({\vec x}-{\vec y} N_2^{-1} (N_1+Q)) \, Q \, ({\vec x}-{\vec y} N_2^{-1} (N_1+Q))^T}
%e^{-{\vec z} \, K \, {\vec z}^T}  \ ,
%\end{equation}
%where $z={\vec x}-{\vec y} \,N_2^{-1} (N_1+K)$. We can now make the replacement
%\begin{equation}
%{\vec v} \cdot {\partial \over \partial {\vec x}} \ \longrightarrow  - \big( {\vec v} \, (N_1+K)^{-1} N_2 \big) \cdot {\partial \over \partial {\vec y}} \ .
%\end{equation}
Fulfilling equation \eqref{hermite} requires ${\vec v}={\vec v}_k$ to be an eigenvector of $(N_1+K)^{-1} N_2$.
While the eigenvalues of $N_1$ and $K$ only depend on $\omega_k$, those of $N_2$ are functions of $p_k$.
The corresponding eigenvalue of the transfer matrix is
\begin{equation}
\Lambda_{k,n}= e^{i n p_k}  \, {\big(1+c^2 e^{-i p_k}\big)^{2 n} \over \big(c \,( a +\omega_k)  \big)^{2 n}}\,  \Lambda_0 \ .
\end{equation}
%and evaluate the integral, obtaining
%\begin{equation}
%{\tilde \Lambda}_0 \, e^{{\vec y} \, Q \, {\vec y}^T}  \, \left( \!  -\big( {\vec v} \, (N_1+Q)^{-1} N_2 \big) \cdot {\partial \over \partial {\vec y}}  \right)^{\!\!n} e^{-{\vec y} \, Q \, {\vec y}^T} \ .
%\end{equation}
%Fulfilling equation \eqref{hermite} requires ${\vec v}={\vec v}_k$ to be an eigenvector of $(N_1+Q)^{-1} N_2$, whose eigenvalues are
%\begin{equation}
%- e^{2 \pi i {k \over L}} \!  \left( { 1+ c^2 e^{- 2 \pi i {k \over L}}  \over c \, (a+ \omega_k )} \right)^{\!\!2}   \ .
%\end{equation}
The ansatz \eqref{fnhermite} can be generalized by allowing for directional derivatives associated to different eigenvectors ${\vec v}_k$.
In this way it can describe states with an arbitrary number of excitations of different momenta.

\subsection{Conserved charges}

The eigenvalue of the transfer matrix on a general state is
\begin{equation}
\Lambda = e^{ i p} \, \prod_{k=0}^{L-1} \, {\big(1+c^2 e^{-i p_k} \big)^{2 n_k} \over \big(c \,( a +\omega_k)  \big)^{2 n_k+1}} \ ,
\end{equation}
where $n_k$ is the number of excitations with momentum $p_k$ and $p=\sum_k p_k \, n_k$. Consistently with \eqref{Tt0} and
\eqref{linearT}, the exponential prefactor on the {\it rhs} is the eigenvalue of the lattice shift operator, $e^{i {\bf P}}$.

We will obtain the complete tower of conserved charges from the expansion of $\log \Lambda$ around $u=0$. The following equality holds (see Appendix C)
\begin{equation}
\log \Lambda =i p \,-\,\sum_{k=0}^{L-1} \! \Big(n_k\!+\!{1 \over 2}\Big)  \log {a+\omega_k \over a- \omega_k} \,+\, \sum_{k=0}^{L-1} \!n_k \log {1+c^2 e^{-i p_k} \over 1+c^2  e^{i p_k}\;\;}% - \log \Big(1\! -\!(- c^2)^L \Big)
\ .
\label{Lsimple}
\end{equation}
The second term on the {\it rhs} only gives rise to odd powers of $u$. The third term, which turns out to be independent of the boson mass, generates even powers. Therefore they contribute to different sets
of conserved charges. A simple basis for the conserved charges can be derived from \eqref{Lsimple}. We will use $c$ as expansion parameter for the
second term and $a^{-1}$ for the third. This just amounts to a linear redefinition of the charges derived using $u$ as expansion parameter. Hence we define
\begin{equation}
\log \Lambda =i p  - \sum_{l=0}^{\infty} a^{-(2l+1)} \langle {\bf Q}_{2l+1} \rangle + i\sum_{l=1}^{\infty} c^{2l} \langle {\bf Q}_{2l} \rangle \ ,
\label{sumQ}
\end{equation}
Equating with \eqref{Lsimple}, we obtain
\begin{equation}
%\langle {\bf Q}_{2l} \rangle= {i \over l } \sum_{k=1}^{L-1} n_k \sin( p_k l) \ , \hspace{.5cm} \langle {\bf Q}_{2l+1} \rangle= {1 \over 2 l\!+\!1 } \sum_{k=1}^{L-1} \Big(n_k\!+\!{1 \over 2}\Big) \,\omega_k^{2l+1}   \ ,
%\langle {\bf Q}_{2l} \rangle= {(-\!1)^l   \over l } \, \sum_{k=1}^{L-1} 2n_k \sin( p_k l) \ , \hspace{.4cm}
\langle {\bf Q}_{2l+1} \rangle= {2 \over 2 l\!+\!1 } \sum_{k=0}^{L-1} \Big(\!n_k\!+\!{1 \over 2}\Big)\,\omega_k^{2l+1}  \ ,\hspace{.4cm}  \langle {\bf Q}_{2l} \rangle= {(-\!1)^l   \over l } \, \sum_{k=1}^{L-1} 2n_k \sin( p_k l)  \ .
\label{vevQ}
\end{equation}
%\barray
%\langle {\bf Q}_{2l} \rangle & =& 2i {(-\!1)^l   \over l } \,\sum_{k=1}^{L-1} n_k \sin( p_k l) \ , \\[2mm]
%\langle {\bf Q}_{2l+1} \rangle &=& {2 \over 2 l\!+\!1 } \, \sum_{k=1}^{L-1} \Big(n_k\!+\!{1 \over 2}\Big)\,\omega_k^{2l+1}   \ ,
%\earray
%There is a total of $L$ conserved charges: $l \leq \left[ {L-1 \over 2} \right]$ ($\left[ {L-2 \over 2} \right]$) for even (odd) charges, where $[\,]$ means retaining the integer part.
%Notice that the even charges are independent of the boson mass.
The first odd charge is ${\bf Q}_1=2{\bf H}$, in agreement with \eqref{linearT}.
The vev's of the conserved charges derived from the transfer matrix expansion are linear in the occupation numbers $n_k$. This is expected since $n_k$ form a basis of conserved quantities of the free boson theory. However, while the occupation numbers can not be derived from a local charge, we will see in the next section that ${\bf Q}_l$ are (quasi) local. Although the sums in \eqref{sumQ} have infinite terms, only a set of $L$ charges can be linearly independent.

%%%%%%%%%%%%%%%%%%%%%%%%%%%%%%%%%%%%%%%%%%%%%%%%%%%%%%%%%%%%%%%%

%\begin{comment}

\section{Quantum inverse scattering method (QISM)}

%An  alternative way to derive the conserved quantities of the free boson lattice model is to apply the QISM,  as has been  done for many  other  integrable  models \cite{F96}.

An alternative way to derive the conserved quantities of an integrable model is to use the QISM \cite{F96}. An advantage of this method is its
applicability to inhomogeneous situations, contrary to the construction \eqref{Tpicture}.  In the previous section we have derived the expectation values of the tower of conserved charges. Here we will use the
QISM to obtain their operator expression.

%\subsection{Operator transfer matrix}

\subsection{Operator form of the $R$ matrices}

%As a preparation we first construct  the operator form of t
The $R$-matrices of the boson model  are maps  from
$\mathbb{R} \otimes \mathbb{R}$ to  $\mathbb{R} \otimes \mathbb{R}$
%that can be written as
%
\beq
R_{x_1 x_2}^{y_1 y_2} = \langle y_1, y_2|  {\bf R}_{12}  |x_1, x_2 \rangle \, ,
\label{a1}
\eeq
with $|x_1, x_2 \rangle$ a complete basis of the Hilbert space on two lattice sites.
%satisfying the orthogonality  conditions
%
%
%\beq
 %\langle y_1, y_2 |x_1, x_2 \rangle = \delta(x_1- y_1) \delta(x_2 - y_2)  \ .
%\label{a2}
%\eeq
We shall  consider instead $\mathbb{R} \otimes \mathbb{R}$ as the real space for the motion of two
particles with coordinates $x_1$ and $x_2$. %,  and conjugate momenta $p_1 =  - i \partial_{x_1}$ and $p_2 = - i \partial_{x_2}$.
We want to express  ${\bf R}_{12}$ as the evolution operator in euclidean time $t$
of a Hamiltonian
${\bf H}_{12}$ acting on the Hilbert space $L_2({\mathbb{R}})  \otimes  L_2(\mathbb{R})$, that is
\beq
{\bf R}_{12} = e^{ -  t \, {\bf H}_{12} } \, .
\label{a3}
\eeq
%This representation will  be  used later  on to construct the  transfer matrix and conserved quantities,  derived in the
%previous section,  using the QISM.
The basic result we use is the euclidean propagator of a harmonic oscillator with mass $m$ and angular frequency $\omega$
\cite{GP}
\barray
G_{m, \omega}(x, x', t) & =  &   \left( \frac{m \omega}{ 2 \pi \sinh( \omega t)} \right)^{1/2}  e^{  -  \frac{ m \omega \{  ( x^2 + x'^2) \cosh(\omega t) - 2 x x' \} }{ 2  \sinh (\omega t) } }  \, .
\label{a5}
\earray
In the limit $\omega \rightarrow 0$, it reduces to the propagator of a free particle of mass $m$.
%\barray
%G_M(R, R', t) & = &  \left( \frac{M}{ 2 \pi  t} \right)^{1/2}  e^{  - \frac{ M (R - R')^2}{ 2 t} }  \ ,
%\label{a4}
%\earray
%and that of a harmonic oscillator with mass $m$ and angular frequency $\omega$
%Eq.\eqref{a4}.

%\subsection{Operator form of the $R$ matrices}

Let us define  the center of mass and relative coordinates  of  the two particle system
\begin{equation}
X =   \frac{x_1 + x_2}{2}, \quad x = x_1 - x_2 \ .  \label{a7}
\end{equation}
%and correspondingly for the pair $(y_1, y_2)$.
The $R$-matrix \eqref{R} factorizes its dependence
on center of mass and relative coordinates as
\begin{equation}
\label{a8}
		R_{x_1 x_2}^{y_1 y_2}
	=
		e^{ -   \bqty{
			\frac{1}{c}   (X - Y)^2
			+   \left( \frac{1}{4 c} + \frac{c}{2} \right)  (x^2 + y^2) -  \frac{1}{2c}   \,   x y
			}
			} \  .
\end{equation}
Comparing with  \eqref{a5}  we make   the identification
\beq
R_{x_1 x_2}^{y_1 y_2} = G_{M,0}(X, Y, 1) G_{m, \omega}(x, y, 1)  \, .
\label{a9}
\eeq
The exponents of both expressions coincide provided that
\beq
\label{a43}
M  = \frac{2}{c} \ ,    \qquad
 m \omega  \coth \omega    = \frac{1}{2c} + c \ , \qquad
 \frac{ m \omega}{\sinh \omega }  =  \frac{1}{2 c}  \, .
\eeq
This also insures agreement between
the normalization factors of the R-matrix and the propagators.
The parameters $m \omega$ and $\omega$ are related  to the variable $u = {\rm arctanh}  (c)$  as
\beq
m \omega  = \frac{1}{ \cos u}\ , \qquad
\omega  =  2 \,  {\rm arsinh}( \tan u)
 \ .
\label{a455}
\eeq
The Hamiltonian ${\bf H}_{12}$,
corresponding  to the Green function \eqref{a9},  is given by the sum of the free particle and harmonic oscillator Hamiltonians
\beq
%{\bf H}_{12}  = \frac{ \hat{p}_X^2}{ 2 M} + \frac{ \hat{p}_x^2}{2 m} + \frac{1}{2}  m \omega^2 x^2   \ ,
{\bf H}_{12}  = \frac{ {\bf p}_X^2}{ 2 M} + \frac{ {\bf p}_x^2}{2 m} + \frac{1}{2}  m \omega^2 {\bf x}^2   \ ,
\label{a16}
\eeq
where %$\hat{p}_X =  \hat{p}_1 + \hat{p}_2$ and  $ \hat{p}_x = (\hat{p}_1 -  \hat{p}_2)/2$
${\bf p}_X =  {\bf p}_1 + {\bf p}_2$ and  ${\bf p}_x = {1 \over 2}({\bf p}_1 -  {\bf p}_2)$
are the center of  mass and relative momenta of the particles. Expanding at small $u$  we obtain
\beq
%{\bf H}_{1,2}  = u  \,  {\bf h}_{1,2} + O(u^3)  , \qquad  {\bf h}_{1,2} =  \frac{\hat{p}_1^2 + \hat{p}_2^2}{2} +   (x_1 - x_2)^2     \ .
{\bf H}_{12}  = u  \,  {\bf h}_{12} + O(u^3)  , \qquad  {\bf h}_{12} =  \frac{{\bf p}_1^2 + {\bf p}_2^2}{2} +   ({\bf x}_1 - {\bf x}_2)^2     \ .
\label{a18c}
\eeq
%and
%
%\beq
%R_{1,2}(u)   = {\bf I}_{1,2} - u  \,  \hat{h}_{1,2}  +  \frac{1}{2} u^2  \hat{h}_{1,2}^{2}     + O(u^3)   \ .
%\label{a18d}
%\eeq
The previous derivation can be repeated  for the massive model. We summarize the main results.
The $R$-matrix \eqref{Rm} corresponds to two harmonic oscillators with masses $M$ and $m$,
and angular frequencies $\Omega$ and $\omega$ given in terms of the  variable  $u$ as
\barray
M \Omega & = &  4 \sqrt{\mu \over \mu_1 } \, \frac{{\rm cn}(u,\mu)}{{\rm dn}(u,\mu)} \ , \qquad \Omega = 2 \   {\rm arcsinh}
\left( \! \sqrt{\mu} \;  \frac{{\rm sn}(u,\mu)}{{\rm dn}(u,\mu)} \right)  \ ,
\label{a23} \\[2mm]
m \omega & = &  \frac{ 1}{ \sqrt{\mu_1} }  \frac{\rm dn(u,\mu)}{\rm cn(u,\mu)} \ ,  \hspace{1.2cm} \omega = 2 \   {\rm arcsinh}
\left(  \frac{\rm sn(u,\mu)}{\rm cn(u,\mu)} \right)  \ .
 \label{a24}
\earray
%\barray
%M \Omega & = &  \frac{ 4 \mu^{1/2} }{ \mu_1^{1/2} } \frac{\rm cn}{\rm dn}, \qquad \Omega = 2 \   {\rm arcsinh}
%\left( \sqrt{\mu}  \frac{{\rm sn}}{{\rm dn}} \right)  \ ,
%\label{a23} \\
%m \omega & = &  \frac{ 1}{ \mu_1^{1/2} }  \frac{\rm dn}{\rm cn},  \qquad \omega = 2 \   {\rm arcsinh}
%\left(  \frac{\rm sn}{\rm cn} \right)  \ .
% \label{a24}
%\earray
%
%where ${\rm cn} = {\rm cn}(u, \mu), \dots$.
%In the massless limit $\mu \rightarrow 0$, one has that  ${\rm cn}, {\rm sn}, {\rm dn} \rightarrow  (\cos , \sin, 1)$, and we recover  \eqref{a455}.
Relations \eqref{a455} are recovered in the massless limit $\mu \rightarrow 0$. The expansion of the  Hamiltonian of the two oscillators
around $u=0$ is
\beq
{\bf H}_{12}  = \sqrt{\mu_1}  u  \, { \bf h}_{12}  + O(u^3) \  , \;\;\;\; {\bf h}_{12} =
% \frac{\hat{p}_1^2 + \hat{p}_2^2}{2} +  \frac{1}{2} \widetilde{m}^2 (x_1^2 + x_2^2) +   (x_1 - x_2)^2    \ .
\frac{{\bf p}_1^2 + {\bf p}_2^2}{2}\,  + \, \widetilde{m}_0^2 \, {{\bf x}_1^2 + {\bf x}_2^2 \over 2} \, +   ({\bf x}_1 - {\bf x}_2)^2    .
\label{a27d}
\eeq
%The expansion of $R_{12}$ is as in \eqref{a18d}.

\subsection{Transfer matrix}

To  implement the  QISM  we use the {\em universal} ${\cal R}$  matrix
defined as  \cite{F96}
\beq
{\cal R}_{12}  = {\bf P}_{12} \, {\bf R}_{12} \, ,
\label{a28}
\eeq
where ${\bf P} _{12}$ is a permutation, such that  in components we have ${\cal R}_{x_1,x_2}^{y_1,y_2} = R_{x_1,x_2}^{y_2,y_1}$.
%
%
%\beq
%{\cal R}_{x_1,x_2}^{y_1,y_2} = R_{x_1,x_2}^{y_2,y_1}  \  .
%\label{a29}
%\eeq
We call ${\cal R}$ {\em universal}  because for the spin models, like the XXZ,
these matrices are derived from a universal ${\cal R}$ matrix  for affine quantum
groups \cite{D86,GR96}.  Using uniformization  variables, the YBE equation expressed in terms of the ${\cal R}$-matrices takes the form
\beq
{\cal R}_{12}(u\!-\!v)  {\cal R}_{13}(u)   {\cal R}_{23}(v)    = {\cal R}_{23}(v)  {\cal R}_{13}(u)   {\cal R}_{12}(u\!-\!v)  \ .
\label{a30}
\eeq
%that is equivalent to (eq. main text).
In the QISM one introduces an auxiliary space $V_a$ and local  quantum spaces $V_j \; (j=1, \dots, L)$,
for the action of the operator
\beq
{\cal R}_{a j}(u) : V_a \otimes V_j \rightarrow  V_a \otimes V_j \ , \quad j=1, \dots, L  \, .
\label{a31}
\eeq
The transfer matrix depicted in \eqref{Tpicture}, can be rewritten in terms of these operators as
%In analogy with \eqref{T}, the transfer operator  ${\bf t}(u)$ is defined as
\beq
{\bf T}(u) = {\rm tr}_a   \big( {\cal R}_{a L}(u) {\cal R}_{a L-1}(u)   \dots {\cal R}_{a 2}(u)  {\cal R}_{a 1}(u)  \big)\ .
\label{a33}
\eeq
In this way we obtain an explicit operator expression for the transfer matrix. We will recover now  the first conserved charges
derived in the previous section from its small $u$  expansion.

%At $u=0$ the R-matrix is the identity and hence ${\cal R}_{1,2}(0) = {\bf P}_{1,2}$.
The permutation operators satisfy %the following properties
${\bf O}_{a k} {\bf P}_{a j} \!= \!{\bf P}_{a j} {\bf O}_{j k}$, for any 2-site operator ${\bf O}$. Applying this property repeatedly
%for ${\bf O}$ being itself a permutation, leads to
%\begin{equation}
%\label{per}
%{\bf O}_{a,k} {\bf P}_{a,j} \!= \!{\bf P}_{a,j} {\bf O}_{j,k} \ , %\quad {\rm tr}_a {\bf P}_{a, L}  = {\bf 1}_L \ ,
%\end{equation}
%for any 2-site operator ${\bf O}$. Using that, we have
%
%\begin{equation}
%{\bf T}(0)\equiv  {\bf T}_0 =   {\bf P}_{1, 2} {\bf P}_{2, 3}   \dots {\bf P}_{L-1, L}  \ ,
%\label{a41}
%\end{equation}
%n agreement with \eqref{Tt0}. The previous property
allows to bring all permutation operators in the transfer matrix to the left of the R-matrices obtaining
%\eeq
%and we can rewrite
\beq
% {\bf T}(u) = {\bf T}_0 \, {\rm tr}_a P_{a, L}  R_{L-1, L}(u)  R_{L-2, L-1}(u)  \dots R_{2,3}(u)  R_{1,2}(u)  R_{a,1}(u)  \ .
 {\bf T} = e^{i{\bf P}} \, {\rm tr}_a \big( {\bf P}_{aL} \, {\bf R}_{L-1, L} \, {\bf R}_{L-2 , L-1} \dots {\bf R}_{23}\,  {\bf R}_{12} \, {\bf R}_{a1}  \big) \ .
 \label{a43}
\end{equation}
%in agreement with \eqref{charges} and \eqref{Tt0}.
%with ${\bf T}_0 =   {\bf P}_{1 2} {\bf P}_{2 3}   \dots {\bf P}_{L-1\, L}$ and which has the same structure as \eqref{Tt0}.
We have dropped the explicit dependence on $u$ in order to simplify the notation.
Observe that the auxiliary space  appears twice on the {\it rhs}, in the operators ${\bf P}_{aL}$ and
 ${\bf R}_{a1}$. To trace over ${V}_a$, we decompose  ${\bf R}_{jk}$ as
 a sum of operators acting in the spaces $V_j$ and $V_k$
\beq
{\bf R}_{jk} = \sum_{\ell} {\bf r}^{(\ell)}_{+j}  \otimes  {\bf r}^{(\ell)}_{-k}    \ .
\label{a44}
\eeq
The transfer matrix is finally given by
\beq
{\bf T} = e^{i{\bf P}}  \sum_{\ell} {\bf r}^{(\ell)}_{+L}  {\bf R}_{L-1 , L}  {\bf R}_{L-2 , L-1}  \dots {\bf R}_{23}  {\bf R}_{12}  {\bf r}^{(\ell)}_{-1} \ .
\label{a45}
\eeq
This expression can also  be written as
\beq
{\bf T} = e^{i{\bf P}}  \sum_{\ell_1, \dots \ell_L} {\bf V}^{\ell_{L}  \ell_{L-1} }_{L}  {\bf V}^{\ell_{L-1} \ell_{L-2} }_{L-1}
\dots {\bf V}^{\ell_{2} \ell_{1} }_{2}  {\bf V}^{\ell_{1} \ell_{L} }_{1}
 \ ,
\label{a45b}
\eeq
with ${\bf V}_j^{\ell \, \ell'} = {\bf r}^{(\ell)}_{+j}  \otimes  {\bf r}^{(\ell')}_{-j}  $, showing explicitly the cyclicity of \eqref{a45}.
%
%\beq
%{\bf V}_j^{\ell \, \ell'} = {\bf r}^{(\ell)}_{+j}  \otimes  {\bf r}^{(\ell')}_{-j}    \ ,
%\label{a44c}
%\eeq
%that  shows explicitly the cyclicity of \eqref{a45}.

%To find the lowest order conserved quantities, we expand  the operator ${\bf R}_{12}$  given in \eqref{a3} with  $t=1$ using \eqref{a27d}.
%Up to second order in $u$,

To find the lowest order conserved quantities we consider the operator ${\bf R}_{12}$  given in \eqref{a3} with  $t=1$, and substitute the expansion
of ${\bf H}_{12}$ derived in the previous subsection.
%it is equivalent to use $u$ or $c$ as expansion parameter.
%From \eqref{a27d} we have
%
%\beq
%{\bf R}_{12} = {\bf I}_{12} - \sqrt{\mu_1} u \, {\bf h}_{12} +  \frac{\mu_1 u^2}{2} \, {\bf h}_{12}^2 + O(u^3)  \ .
%\label{a46}
%\eeq
%Substituting it into \eqref{a45} gives
%Substituting into \eqref{a45} gives
Plugging the expression so obtained into \eqref{a45} gives
\beq
%c \sum_{j=1}^L {\bf h}_{j, j+1} + \frac{c^2}{2}  \left(  \sum_{j=1}^L   {\bf h}_{j, j+1} \right)^2 \!\! - \, \frac{c^2}{2} \, \sum_{j=1}^L  \big[ {\bf h}_{j, j+1}, {\bf h}_{j+1, j+2} \big]  \ .
%{\bf T} = e^{i{\bf P}}  \Big( {\bf I} -    2 \sqrt{\mu_1} u \, {\bf H} + 2 \mu_1 u^2 ({\bf H}^2 -i {\bf Q}_2) + {\cal O}(u^3) \Big) \ .
\log {\bf T} = i{\bf P}  -    2 \sqrt{\mu_1} u \, {\bf H} +  i \mu_1 u^2 \, {\bf Q}_2 + {\cal O}(u^3)  \ .
\label{a47}
\eeq
Identifying ${\bf p}_i=a_x {\boldsymbol \pi}_i$, we recover the Hamiltonian of the lattice model
%This reproduces  \eqref{Texpand} since
%\beq
%{\bf T}_0^{-1}  {\bf T}= {\bf 1} -   2 c {\bf H}+ 2 c^2 ({\bf H}^2  - i  {\bf Q}_2)   \ ,
%\label{a48}
%\eeq
%
%where
%
\beq
{\bf H} = \frac{1}{2}  \sum_{i=1}^L {\bf h}_{i, i+1}= \frac{1}{2} \sum_{i= 1}^L  \left(  {\bf p}_i^2   + ({\bf x}_i - {\bf x}_{i+1})^2 + \widetilde{m}_0^2 \,  {\bf x}_i^2 \right)  \ .
\label{a49}
\eeq
The charge ${\bf Q}_2$ is given by
\beq
{\bf Q}_2 = \frac{i}{2}   \sum_{i=1}^L  \big[ {\bf h}_{i, i+1},{\bf h}_{i+1, i+2} \big]   =  \sum_{i=1}^L  {\bf p}_i ( {\bf x}_{i-1} - {\bf x}_{i+1} )  \ .
\label{a50}
\eeq
%provides the operator form of the conserved quantity whose expectation value is given by \eqref{Q2}.
%It comes from the commutator  in \eqref{a47}. The analogue of that term in the spin 1/2 Heisenberg model
%is given by $\sum_j \vec{S}_j \cdot  ( \vec{S}_{j+1} \times  \vec{S}_{j+2})$ that of course commutes with the
%Hamiltonian $\sum_j \vec{S}_j \cdot \vec{S}_{j+1}$ that breaks time reversal \cite{pink}.
Using this insight, we propose the following operator expression for the even conserved charges of the discretized boson theory
\begin{equation}
%{\bf Q}_0= {\bf P}  \; , \hspace{.8cm} {\bf Q}_{2l}= {(-\!1)^l \over l } \sum_{i=1}^L  a_x {\boldsymbol \pi}_i ( {\bf x}_{i-l}-{\bf x}_{i+l}) \ ,
%{\bf Q}_0= {\bf P}  \; , \hspace{.8cm}
{\bf Q}_{2l}= {(-\!1)^{l\!-\!1} \over l } \, a_x \, {\vec {\boldsymbol \pi}}  \left( S^l -S^{l  T}  \right)  {\vec {\bf x}} \ ,
\label{Qeven}
\end{equation}
%where $l<\left[ {L-1 \over 2} \right]$. The operator form of the odd charges is
with $l \geq 1$. For the odd charges we have
\begin{equation}
%{\bf Q}_{2l+1}= {1 \over 2l\!+\!1} \sum_{i,j=1}^L \Big( a_x^2   {\boldsymbol \pi}_i \, (K^{2l})_{ij} \, {\boldsymbol \pi}_j + {\bf x}_i \, (K^{2l+2})_{ij} \, {\bf x}_j \Big) \ ,
{\bf Q}_{2l+1}= {1 \over 2l\!+\!1} \Big( a_x^2 \, {\vec {\boldsymbol \pi}} \, K^{2l} \, {\vec {\boldsymbol \pi}} + {\vec {\bf x}} \, K^{2l+2} \, {\vec {\bf x}} \Big) \ ,
\label{Qodd}
\end{equation}
with $l \geq 0$ and $K^2$ given in \eqref{Q}. %The lowest odd charge ${\bf Q}_1$ is the Hamiltonian of the lattice model.
In Appendix C we show that these charges indeed commute and their expectation values agree with \eqref{vevQ}.

Using that $S^L=1$, it is immediate to see that there are $\left[ {L \over 2} \right]+1$  linearly independent odd charges \eqref{Qodd} and
$\left[ {L-1 \over 2} \right]$ even charges \eqref{Qeven}, where $[\cdot]$ denotes retaining the integer part. This makes a set $L$ quasi local operators
that, avoiding ${\bf P}$, can be used as a basis of the conserved charges of the discretized bosonic theory.
The higher the charge, the farther the sites it couples. Even charges couples sites $i,i\pm l$ and
odd charges couples sites up to $i,i\pm(l+1)$. In the continuum limit, long range effects translate into multiple spatial derivatives.

We observe  that the operator ${\bf Q}_2$ provides a method to compute the
momentum of the bosonic excitations,  that of course agrees with the one obtained using the lattice shift operator $ e^{ i {\bf P}}$.
However, unlike $e^{i {\bf P}}$, the operator ${\bf Q}_2$ has the advantage of being local.
Finally, it is worth noticing  that
for a massless bosonic theory in the continuum limit
 \barray
 {\bf H} & \rightarrow   &   \frac{1}{2}  \int dx \left( ( \partial _t \phi)^2 + ( \partial _x \phi)^2  \right) =  \frac{2 \pi}{L}  (L_0 + \bar{L}_0 )  \ , \label{a53} \\
 {\bf Q}_2 & \rightarrow   &  \int dx \,  \partial _t  \phi   \partial _x \phi =  \frac{2 \pi}{L} (L_0 -  \bar{L}_0)   \ .
 \label{a53b}
 \earray
 where $L_0$ and $\bar{L}_0$ are the holomorphic and antiholomorphic Virasoro operators
 of the $c=1$ CFT of the massless boson \cite{CFT}.
 These equations % \eqref{a53} and \eqref{a53b}
 imply  that  $H + Q_2$ and $H -  Q_2$ provide a local
 lattice version of the Virasoro operators $L_0$ and $\bar{L}_0$.

%%%%%%%%%%%%%%%%%%%%%%%%%%%%%%%%%%%%%%%%%%%%%%%%%%%%%%%%%%%%%%%%%%%%%%%
\section{Scattering $S$ matrix}
%%%%%%%%%%%%%%%%%%%%%%%%%%%%%%%%%%%%%%%%%%%%%%%%%%%%%%%%%%%%%%%%%%%%%%%

The $R$ matrices introduced in previous sections are inhomogeneous deformations
of the Boltzmann weights  of a massless and massive boson in a square lattice.
A Wick rotation of these weights amounts  to study the  discrete  time evolution of the
boson. This is obtained  replacing $c \in \mathbb{R}$ by  $i c$ in \eqref{R},
such that ${\bf R}$ becomes essentially  a pure phase.
In this section we adopt a different point of view
demonstrating
that the Wick rotated Boltzmann weights can be understood  as relativistic scattering
$S$-matrices.
%This result may seem as expected since the boson model in the lattice  is a free theory
%\textcolor{red}{why?}.
This result  is far from obvious but, as mentioned  in the introduction,  there are
examples where this occurs,
 as the 6 vertex and the sine-Gordon model,  along  with their elliptic
deformations. An additional ingredient of our construction  is that the $S$-matrix
describes the scattering of massive particles with continuous degrees of freedom.
%\textcolor{red}{ although there are some Statistical Mechanical models with property. }

%Secondly, the degrees of freedom involved in the scattering are continuous variables, not discrete ones
%with a topological meaning  as occurs in quantum field theory models like sine-Gordon.
%This means that the existence of a $S$-matrix constructed out  from the Boltzmann weights of the boson is not guaranteed at all.
%\textcolor{red}{why we think that such S-matrix interpertation exists?}

Let us briefly review the factorized $S$-matrix theory.
This theory describes the scattering of a set of a particles $\{ A_i \}_{i=1}^N$
in a relativistic quantum field theory \cite{ZZ79,GR96,M10}. If  the particles are massive, their
energy and momentum are parameterized in terms of their rapidity $\theta \in \mathbb{R}$  as $(p^0_i, p^1_i)= m_i (\cosh \theta, \sinh \theta)$,
where $m_i$ is the mass of the particle $A_i$.
%\textcolor{red}{xxx} If the theory is massless, the right/left  moving particles
%are parameterized by $p^0_i =  \pm p^1_i = \Lambda e^{ \pm \theta}$, where $\Lambda$ is a energy scale.
The  two particle scattering process between incoming and outgoing  asymptotic states
is given by
\beq
{\bf S} |A_i(\theta_1) , A_j(\theta_2) \rangle_{\rm in} = \sum_{j k}  S_{ij}^{k \ell}(\theta_{12}) |A_k(\theta_2) , A_\ell(\theta_1) \rangle_{\rm out}
\label{s0}
\eeq
where, by relativistic invariance,  the matrix $S$ only depends on the difference of rapidities  $\theta_{12} = \theta_1\! -\! \theta_2 >0$.
Factorization guarantees that the two  particle amplitude \eqref{s0} completely determines all possible scattering processes. It requires for consistency
\beq
 \sum_{p_i} S_{i_1 i_2}^{p_1 p_2}(\theta_{12} ) S_{p_2 i_3}^{p_3 j_3}(\theta_{13} )
 S_{p_1 p_3}^{j_1 j_2}(\theta_{23}) =
  \sum_{p_i} S_{i_2 i_3}^{p_2 p_3}(\theta_{23}) S_{i_1 p_2}^{j_1 p_1}(\theta_{13}) S_{p_1 p_3}^{j_2 j_3}(\theta_{12}) \ ,
\label{fact}
 \eeq
which is equivalent to the YBE. %This implies that the amplitude $S_{ij}^{kl}(\theta_{12}$ completely characterizes the scattering theory.

Rapidities can be allowed to take complex values within the so called physical strip,  ${\rm Im} \, \theta \in (0, \pi)$.
%\textcolor{red}{(this is the definition of the physical strip used by Zamos and Mussardo)}.
With this extension $\theta$ uniformizes the branch cuts both in the $s$ and $t$-scattering channels, which map respectively to ${\rm Im} \, \theta =0$ and
${\rm Im} \, \theta =\pi$. Hence the amplitude  $S_{ij}^{kl}(\theta)$ must  be a  meromorphic function
whose only singularities are poles in the imaginary axis of the physical strip, associated to the eventual  appearance of a bound state
 \cite{M10}.
%Equation \eqref{s0} can be given an algebraic form due to  Faddeev and Zamolodchikov  in terms of an operator ${\bf A}_i(\theta)$ whose  action  on the Hilbert space vacuum $|0 \rangle$ creates a particle in the state $| A_i(\theta) \rangle$,
%
%\beq
%A_i(\theta) |0 \rangle =
%| A_i(\theta) \rangle   \, .
%\label{s0a}
%\eeq
%Equation \eqref{s0} is  equivalent to the exchange  relation
%
%\beq
%{\bf A}_i(\theta_1)  {\bf A}_j(\theta_2)  = \sum_{j k}  S_{ij}^{k \ell}(\theta_{12})  {\bf A}_k(\theta_2)  {\bf A}_\ell(\theta_1)  \ .
%\label{s0b}
%\eeq
%
Besides \eqref{fact}, the  $S$-matrix has to fulfill the following axioms
\begin{eqnarray}
\hspace{-1.5cm} {\rm   (i) \; Normalization:}   \hspace{2cm} &&  \lim_{\theta  \rightarrow 0}  {\bf S}(\theta)  = {\bf 1} \label{S0} \\[4mm]
\hspace{-1.5cm}  {\rm    (ii)\;  Unitarity: } \hspace{3cm} && {\bf S}(\theta)   {\bf S}(-\theta) = {\bf 1}  \\[5mm]
\hspace{-1.5cm} {\rm  (iii) \; Real \; analiticity:  } \hspace{1.9cm}   && {\bf S}^\dagger(\theta)  =  {\bf S}(-\theta^*)  \\[5mm]
\hspace{-1.5cm} {\rm    (iv)\;  Crossing \; symmetry:}   \hspace{1.1cm} &&  S_{ij}^{k \ell}(\theta) =   S_{j \bar{\ell}}^{\bar{i} k}(i \pi - \theta)
%& {\rm    (v)\;  Factorization:}   \hspace{2cm} &
\end{eqnarray}
%\beq
% \lim_{\theta  \rightarrow 0}  S_{ij}^{k \ell}(\theta) = \delta_{i}^k \delta_{j}^{\ell}
%  \Longleftrightarrow  \lim_{\theta  \rightarrow 0}  {\bf S}(\theta)  = {\bf 1}  \, .
%\label{s1}
% \eeq
  %\beq
% \sum_{j_1, j_2}   S_{j_1 j_2}^{i_1 i_2}(\theta)  S_{k_1 k_2}^{j_1 j_2}( -\theta) = \delta_{k_1}^{i_1}  \delta_{k_2}^{i_2}
%   \Longleftrightarrow  {\bf S}(\theta)   {\bf S}(-\theta) = {\bf 1}  \, .
% \label{s2}
% \eeq
%\beq
%  {\bf S}^\dagger(\theta)  =  {\bf S}(-\theta^*) \, .
% \label{s3}
% \eeq
%\beq
 % S_{ij}^{k \ell}(\theta) =   S_{j \bar{\ell}}^{\bar{i} k}(i \pi - \theta) \ .
% \label{s4}
% \eeq
%  \beq
% \sum_{p_1 p_2 p_3} S_{i_1 i_2}^{p_1 p_2}(\theta_{12} ) S_{p_2 i_3}^{p_3 j_3}(\theta_{13} )
% S_{p_1 p_3}^{j_1 j_2}(\theta_{23}) =
%  \sum_{p_1 p_2 p_3} S_{i_2 i_3}^{p_2 p_3}(\theta_{23}) S_{i_1 p_2}^{j_1 p_1}(\theta_{13}) S_{p_1 p_3}^{j_2 j_3}(\theta_{12}) \ .
%\nonumber
% \eeq
 Condition (i) means that no scattering takes place when the relative velocities of the two  particles vanishes.
 Condition (ii) is obtained applying \eqref{s0} twice. Conditions (ii) and (iii) imply physical unitarity
 ${\bf S}^\dagger(\theta)  {\bf S}(\theta) = {\bf 1}$. Condition (iv) relates the scattering channel $A_i \times A_j \rightarrow A_k \times A_\ell$
 to the crossed channel $A_j  \times A_{\bar{\ell}}  \rightarrow A_{\bar{i}}  \times A_k$, where the bar denotes the corresponding antiparticle.
 %Finally, condition (v) guarantees the associativity of the algebraic relation \eqref{s0b} and is equivalent to the YBE.
 % is a generalized ``Jacobi identity''  that guarantees a consistent  factorization of the scattering between three
 %particles,  with rapidities $\theta_1 > \theta_2 > \theta_3$,  as products of three  two particles scatterings, namely:
 %$123 \rightarrow 213 \rightarrow 231 \rightarrow 321$  equal to  $123 \rightarrow 132  \rightarrow 312  \rightarrow 312$.

%The amplitude  $S_{ij}^{kl}(\theta)$ must  be a  meromorphic function in the complex $\theta$ plane
%physical strip ${\rm Im} \; \theta \in (0, \pi)$
%whose only singularities are poles in the imaginary axis associated to the eventual  appearance of a bound state  \cite{M10}.

\subsection{The trigonometric  $S$ matrix }

We make the following change of variables from the parameter $u$ employed in the massless Stat. Mech. model
to a rapidity variable $\theta$, which implements the Wick rotation of the Boltzmann weights
\beq
u = \, \frac{ \theta}{2 i} \qquad   \Longrightarrow \qquad c(\theta) =  -  i \tanh \frac{ \theta}{2} \ .
\label{s5}
\eeq
%which implements the Wick rotation of the Boltzmann weights.
%such that
%
%\beq
%c(\theta) =  i \tanh \frac{ \theta}{2} \ .
%\label{s6}
%\eeq
%\textcolor{red}{$\theta$ parameterizes the energy-momentum of the particles
%which we assume have the same mass $(p^0, p^1)  = m (\cosh \theta, \sinh \theta)$}.
Replacing this into \eqref{R}, and allowing for a $\theta$-dependent proportionality constant,
we define ${\bf S}(\theta)=g(\theta) {\bf R}(\theta)$ with
 %
% \begin{equation}
%\label{s7}
%S_{x_1,x_2}^{y_1,y_2}(\theta)
%=   g(\theta)  R_{x_1,x_2}^{y_1,y_2}(c(\theta))
%	\quad x_1, x_2, y_1, y_2 \in \mathbb{R} \  ,
%\end{equation}
%\begin{equation}
%		R_{x_1,x_2}^{y_1,y_2}(c(\theta) )
%	= \frac{ \coth \frac{\theta}{2} }{2 \pi  i }     \
%		e^{ {i \over 2} \bqty{
%			 \coth \! \frac{\theta}{2}   (x_1 - y_1)^2
%			+  \coth \!\frac{\theta}{2}  (x_2 - y_2)^2
%			-  \tanh \!\frac{\theta}{2}  (x_1 - x_2)^2
%			-  \tanh \! \frac{\theta}{2}  (y_1 - y_2)^2
%			}
%			} \  ,
%	\quad x_1, x_2, y_1, y_2 \in \mathbb{R} \  ,
%\end{equation}
\begin{equation}
		R_{x_1 x_2}^{y_1 y_2}(\theta )
	= \frac{  i \coth \frac{\theta}{2} }{2 \pi  }     \
		e^{ - {i \over 2}
			 \coth \! \frac{\theta}{2}   \bqty{ (x_1 - y_1)^2
			+  (x_2 - y_2)^2 }
			+    {i \over 2} \tanh \!\frac{\theta}{2}  \bqty{ (x_1 - x_2)^2
			+ (y_1 - y_2)^2
			}
			} \  .
\label{Rtheta}
%	\quad x_1, x_2, y_1, y_2 \in \mathbb{R} \  ,
\end{equation}
This insures that the factorization condition \eqref{fact} is satisfied.
We want to stress the very different meaning of $u$ in the Stat. Mech. model, where it parameterizes the lattice anisotropy, and $\theta$
in the scattering theory, which is a dynamical variable parameterizing the  dispersion relations
$(p^0, p^1)  = m (\cosh \theta, \sinh \theta)$.

The continuous nature of field variables in the boson model implies the presence of an infinite set of particles in the scattering theory,
labelled by a continuum index $x$.
The R-matrix \eqref{Rtheta} is easily seen to be compatible with the axioms (i)-(iv), translating them into conditions on the function $g$.
%These conditions, together with certain analyticity properties allow to determine $g$.
%and the  function $g(\theta)$ will fixed by imposing  conditions  (i)-(v) and certain analiticity   properties.
Using \eqref{Id}, normalization (i) holds provided
\beq
\lim_{\theta \rightarrow 0}  g(\theta) = 1 \ .
\label{s16}
\eeq
The unitarity condition (ii)  reads
\beq
 \int \! d {\vec y} \,
 S_{y_1 y_2}^{z_1 z_2}(\theta)   S_{x_1 x_2}^{y_1 y_2}(-\theta)  = \delta(x_1\! -\! z_1) \delta(x_2 \!-\! z_2)  \hspace{4mm}  \Longrightarrow  \hspace{4mm}
 g(\theta) g(- \theta) = 1 \ .
 \label{s19}
 \eeq
%which is satisfied if
%
%\beq
%g(\theta) g(- \theta) = 1 \ .
%\label{s21}
%\eeq
The integral is done assuming that $\theta$ is real.
The latter assumption also guarantees the convergence of the integrals in \eqref{fact}.
Real analiticity (iii) implies
\beq
\left( S^{x_1 x_2}_{y_1 y_2}(\theta) \right)^*   = S_{x_1 x_2}^{y_1 y_2}(- \theta^*) \hspace{4mm}  \Longrightarrow  \hspace{4mm} g^*(\theta) = g( - \theta^*)  \ .
\label{s8}
\eeq
Crossing symmetry has been used to determine the proportionality constant in the linear relation between $u$ and $\theta$ \eqref{s5}. Then (iv) holds if
%Crossing symmetry (iv) holds if
%
\beq
 S_{x_1 x_2}^{y_1 y_2}(\theta)    = S_{x_2 y_2}^{x_1 y_1}(i \pi -  \theta)  \hspace{4mm}  \Longrightarrow  \hspace{4mm} g(i \pi \!-\! \theta) = - g(\theta)  \coth^2 \frac{ \theta}{2}  \ .
\label{s10}
\eeq
Notice that the counterpart of the physical strip in the Stat. Mech. model, is the requirement that the real part of $u$ takes values in the interval
$(0, {\pi \over 2})$. We have already encountered this restriction when requiring a regular large field limit of the R-matrix \eqref{R}. Hence the natural domains of dependence of the variables $\theta$ and $u$ map to each other, supporting the $S$-matrix interpretation of the discretized free boson.
%which holds if
%
%\beq
%g^*(\theta) = g( - \theta^*)  \ .
%\label{s9}
%\eeq
%
%which holds if
%
%\beq
%g(i \pi - \theta) = - g(\theta)  \coth^2 \frac{ \theta}{2}  \ ,
%\label{s18}
%\eeq
%where we have used that $\coth \frac{ i \pi -\theta}{2} = - \tanh \frac{\theta}{2}$.
%
%This equation uses  the limit
%
%\beq
%\delta(x-y) =
%\lim_{t \rightarrow 0}  \frac{1}{ \sqrt{ 2 \pi i t} } e^{ \frac{i}{2} \frac{ (x-y)^2}{t} } \ .
%\label{s14}
%\eeq
%
%The unitarity condition (ii)  reads
%
%\beq
% \int d y_1 d y_2
% S_{y_1,y_2}^{z_1,z_2}(\theta)   S_{x_1,x_2}^{y_1,y_2}(-\theta)  = \delta(x_1 - z_1) \delta(x_2 - z_2) \ ,
% \label{s19}
% \eeq
%which is satisfied if
%
%\beq
%g(\theta) g(- \theta) = 1 \ .
%\label{s21}
%\eeq
%The integrals in  \eqref{s19} are done assuming that $\theta$ is real.
%The latter assumption also guarantees the convergence of the integrals in  (v),
%a condition that is verified since it is just  the YBE of the $R$ matrix \eqref{R}.
%The equations satisfied by $g(\theta)$ take the form
%
  %
% \barray
% g(0) =  1,  &  & \quad   g^*(z) = g(z^*)  \ ,  \label{m7} \\
%g(z) g(- z)   =   1,  & &  \quad g\left(  \frac{1}{2} - z \right)  =     g(z)  \cot^2(\pi z)    \ ,
%\nonumber
%\earray
%in terms of the variable
%
%\beq
% z= \frac{  \theta}{2 \pi i} \ .
% \label{m6}
% \eeq

Using the infinite product decomposition
\beq
%\cot( \pi z) = - \prod_{n = - \infty}^\infty \frac{ n + \frac{1}{2} + z}{ n -z}  \ ,
\coth {\theta \over 2} = i \prod_{n = - \infty}^\infty \frac{ n + \frac{1}{2} + {\theta \over 2 \pi i}}{ n -{\theta \over 2 \pi i}}  \ ,
\label{m16}
\eeq
we find the following  solution of equations  \eqref{s16}-\eqref{s10}  (see  Appendix D)
\beq
%g(z) = \lim_{M \rightarrow \infty}
% \prod_{n=- M}^M  \left(   \frac{  \Gamma( n + z ) \Gamma \left( n +  \frac{1}{2} - z \right)     }{   \Gamma( n - z )  \Gamma \left( n +  \frac{1}{2}  + z \right) } \right)^2
g(\theta) = \lim_{M \rightarrow \infty}
 \prod_{n=- M}^M  \left(   \frac{  \Gamma \left( n + {\theta \over 2 \pi i} \right) \Gamma \left( n +  \frac{1}{2} - {\theta \over 2 \pi i}  \right)     }{   \Gamma \left( n - {\theta \over 2 \pi i}  \right)  \Gamma \left( n +  \frac{1}{2}  + {\theta \over 2 \pi i}  \right) } \right)^2
\ .
\label{m20}
\eeq
The parameter $M$ and its limit is required for regularization.
%\textcolor{red}{why $\lim_{M \to 0}$}.
This is a meromorphic function with  double poles at %$z= \frac{ 1}{2}, \frac{3}{2}, \dots$
$\theta= \pi i,  3 \pi i, \dots$
and quadrupole poles at $\theta = -2 \pi i,-4 \pi i, \dots$.
Hence  there are not singularities on the physical strip except at its boundary $\theta  = i \pi$.
This double pole of $g$ conspires  with the simple zero of $\coth \frac{\theta}{2}$ in \eqref{Rtheta},
and the exponential term to give
%At that point,  the factor $\cot \frac{\theta}{2}$ of \eqref{Rtheta}  has a simple zero, that combines with  the exponential term to give
%$S_{x_1 x_2}^{y_1 y_2} (i \pi) = \delta(x_1 - x_2) \delta(y_1 - y_2)$
\begin{equation}
S_{x_1 x_2}^{y_1 y_2} (i \pi) = \delta(x_1 - x_2) \delta(y_1 - y_2) \ .
\end{equation}
%that also follows from equation ${\bf S}(0) = {\bf 1}$ using  crossing symmetry.
This equation also  follows from the normalization condition ${\bf S}(0) = {\bf 1}$ and crossing symmetry.
\begin{figure}[t!]
\vspace{.2cm}
\begin{center}
\includegraphics[width=0.4\textwidth]{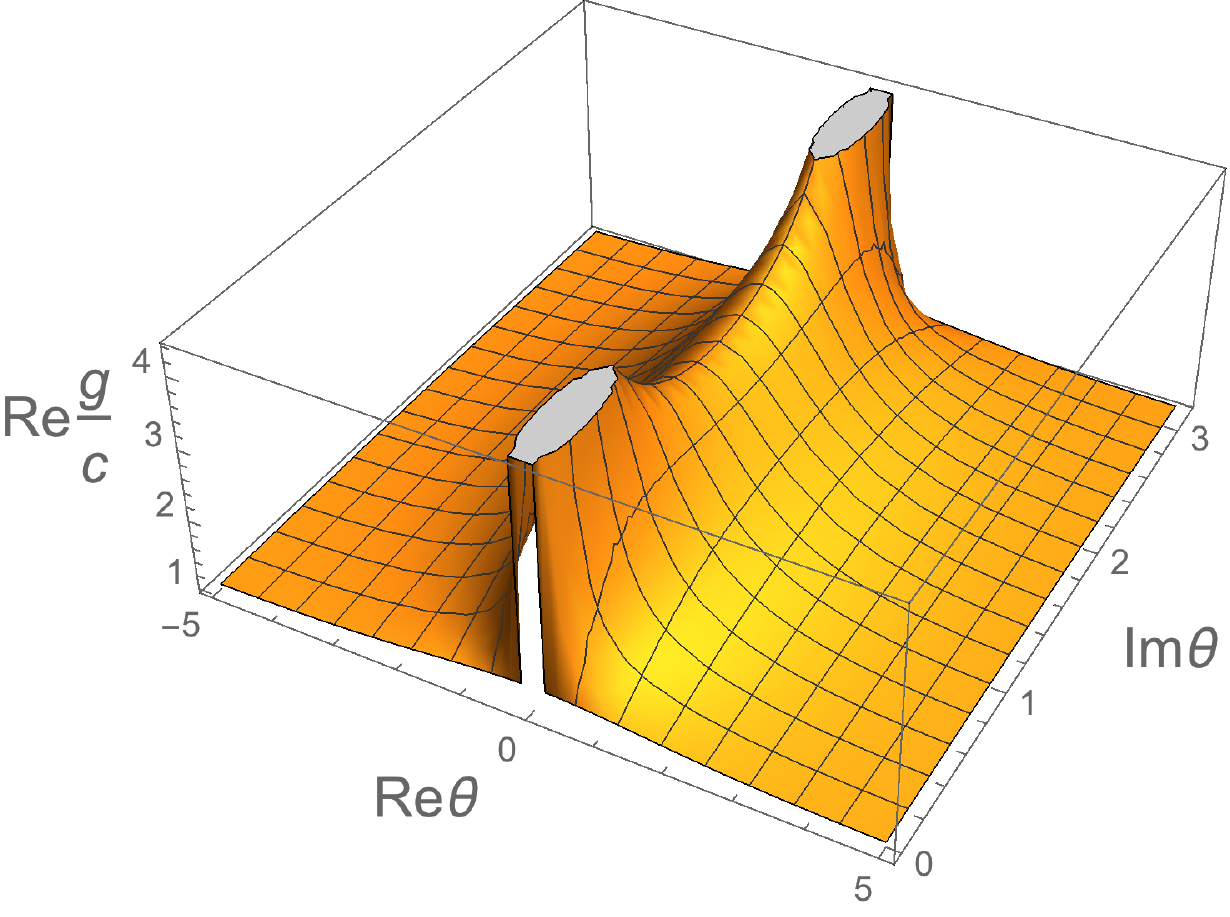} \qquad
\includegraphics[width=0.4\textwidth]{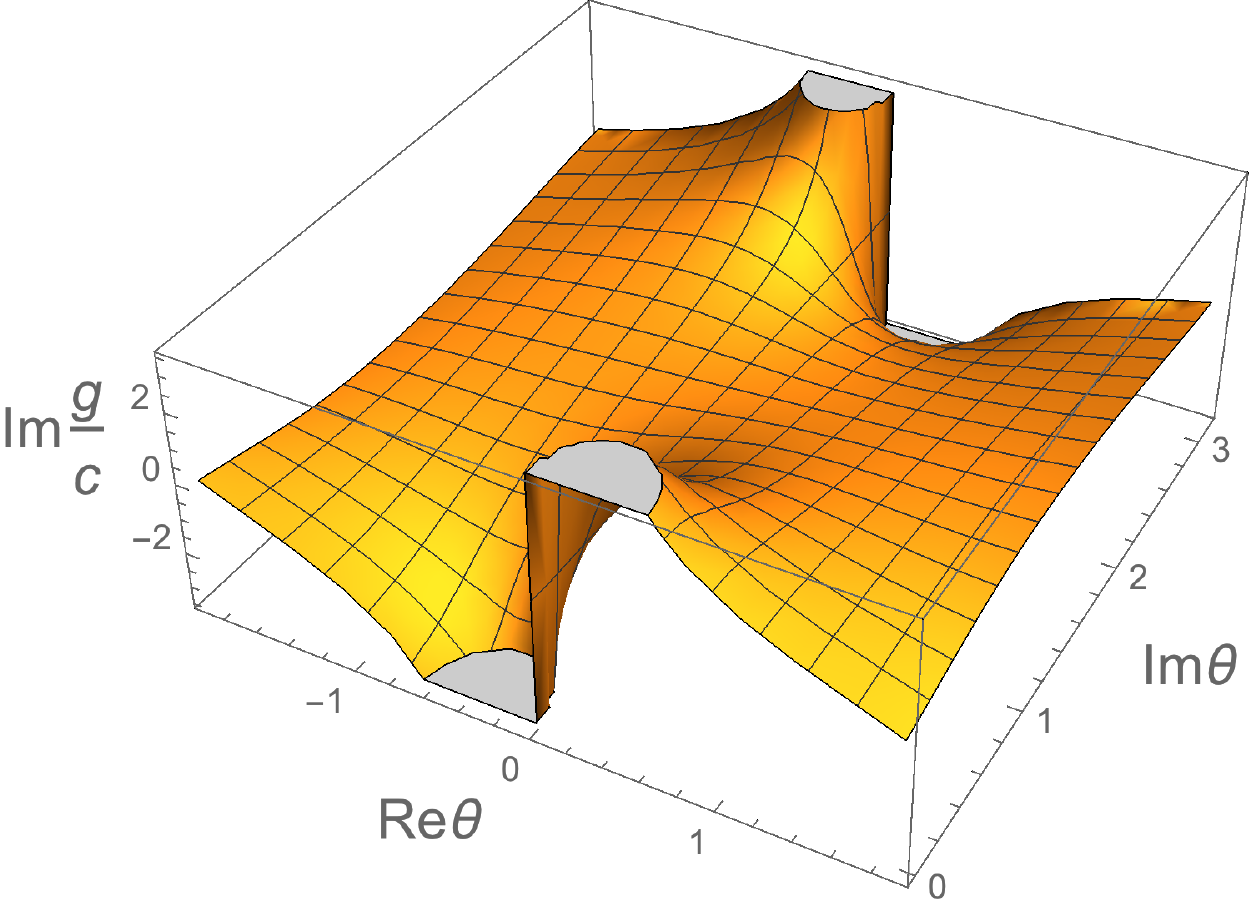}
\includegraphics[width=0.5\textwidth]{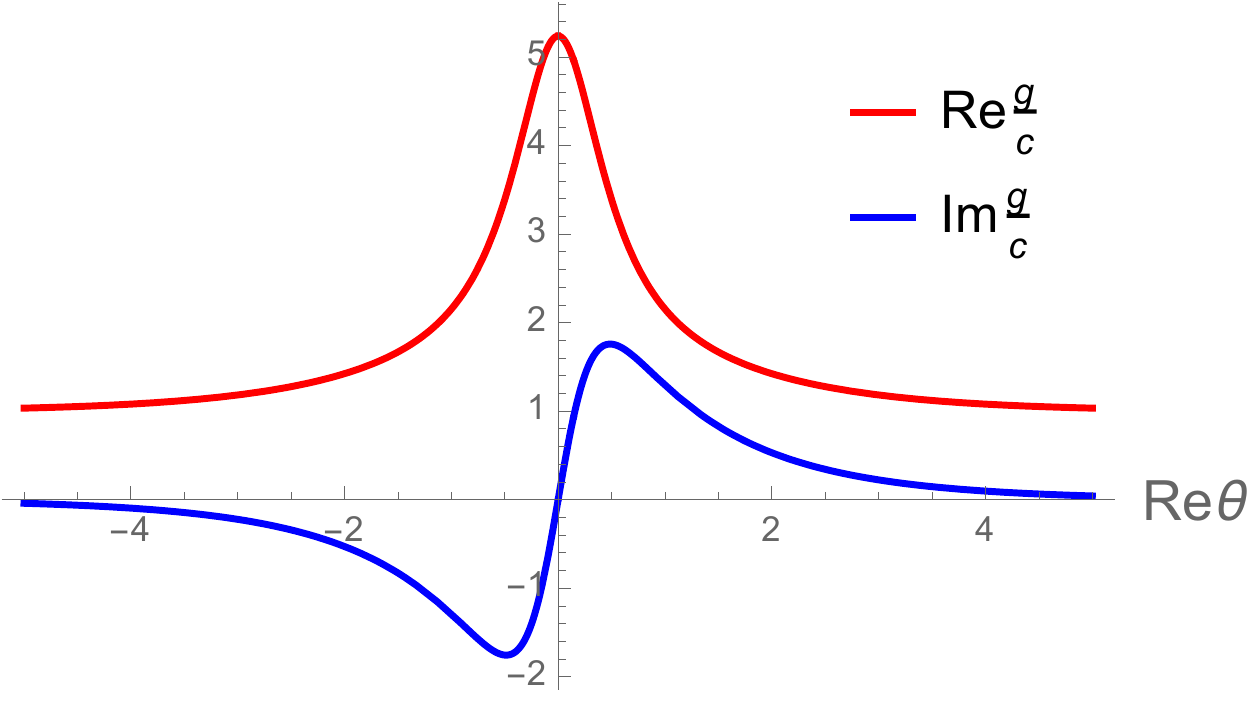}
\caption{
Above: 3-dimensional plot of the real and imaginary parts of ${g \over c}=i g \coth{\theta \over 2}$ in the physical strip. Below: Section of the previous plots at ${\rm Im}(\theta)={\pi \over 6}$.  }
\label{g}
\end{center}
\end{figure}
%\textcolor{red}{meaning}.
An integral representation of the $g$ function in the region  $|{\rm Im} \; \theta| < \pi$ is given by (see Appendix D)
%region $|{\rm Re} \; z| < \frac{1}{2}$, that  corresponds to $|{\rm Im} \; \theta| < \pi$, is given by
%
\begin{equation}
g(\theta)= {\rm exp} \left( -i  \int_0^\infty \frac{dt}{t}  { \sin( 2 \theta t ) \over \cosh^2(\pi t)}
\right)  \ .
\label{m21}
\end{equation}
This expression can be explicitly integrated, obtaining
\begin{equation}
g(\theta) =i {\rm exp} \left[ -{2 i \over \pi} \left( \theta \log {1 -e^\theta \over 1+ e^\theta} +{\rm Li}_2 (e^\theta)-{\rm Li}_2 (-e^\theta) \right) \right] \ ,
\label{poly}
\end{equation}
where ${\rm Li}_2(z)$ is the polylogarithmic function. The product $i g \coth{\theta \over 2}$ has been plotted in Fig.\ref{g}.
This graphic renders crossing evident, since at the level of that function it just amounts symmetry under the exchange $i \pi -\theta \leftrightarrow \theta$.
Finally,  we should mention that $g$ is not unique due to the so called  CDD ambiguity.
%The $g$ function is not unique due to the so called  CDD ambiguity, which means that the scattering
Namely, the scattering matrix can be multiplied by a meromorphic function satisfying \eqref{s16}-\eqref{s8} and invariant under crossing, which adds extra zeros and poles
\cite{ZZ79}. What we have obtained above is a minimal solution that  does not contain CDD poles.

%\textcolor{red}{say more: ambiguity in $g$, physical implications of $g$}
%\barray
%g(z) & =  &  {\rm exp} \left( \int_0^\infty \frac{dt}{t}  \frac{ \sinh( tz)}{ \cosh^2(t/4)} \right)
%\label{m21}  \\
%& =  &  {\rm exp} \left[ \frac{ i \pi}{2} + 4 z \log \frac{ 1 - e^{ 2 \pi i z}}{ 1 + e^{ 2 \pi  i z}} + \frac{ 2 i}{\pi} \left( {\rm Li}_2( -  e^{ 2 \pi i z}) - {\rm Li}_2(  e^{ 2 \pi i z} )  \right) \right]
%\label{m21b}
%\earray
%where ${\rm Li}_2(z)$ is the polylogarithmic function.
%

\subsection{The elliptic  $S$-matrix }

Guided again by crossing we choose
\beq
u = \frac{ K \,  \theta}{\pi i} \ ,
\label{s25}
\eeq
with $K \equiv K(\mu)$ defined in \eqref{K}. Substituting this change of variables in \eqref{cumm} leads to
\beq
c(\theta, \mu) = \sqrt{\mu_1} \frac{ {\rm sn} \! \left(  \frac{ K  \,  \theta}{\pi i}, \mu   \right) }{
{\rm cn} \! \left(  \frac{ K  \,  \theta}{\pi i}, \mu   \right) {\rm dn} \! \left(  \frac{ K  \,  \theta}{\pi i}, \mu  \right)} , \quad
{\widetilde m}(\theta, \mu) =  \sqrt{\frac{4 \mu} {\mu_1}} \,  {\rm cn} \! \left( \! \frac{ K  \,  \theta}{\pi i}, \mu  \! \right) \ .
\label{cu}
\eeq
%which in the massless limit  becomes \eqref{s5}.
It is easy to show that  $c(\theta, \mu)$ becomes purely imaginary and  ${\widetilde m}(\theta, \mu)$  real,   when $\theta \in \mathbb{R}$.
%It is easy to show that these functions becomes purely imaginary  when $\theta$ is real. }
%$c(\theta, \mu)$ and $\nu(\theta, \mu)$, obtained by substituting the change of variables above in \eqref{cumm},
%become purely imaginary when $\theta$ is real.
We define again ${\bf S}(\theta,\mu)=g(\theta,\mu) {\bf R}(\theta,\mu)$. As in the previous section, the function $g$ will be determined by
the $S$-matrix theory axioms. Equations \eqref{s16}-\eqref{s8} remain unaltered in the elliptic  case. Equation \eqref{s10}, derived from crossing, also applies
after replacing
$\coth^2(\theta/2)$ by $-1/c^2(\theta, \mu)$.
%Similarly to \eqref{s7} we define
 %
% \begin{equation}
%\label{s57}
%\tilde{S}_{x_1,x_2}^{y_1,y_2}(\theta, \mu)
%=   \tilde{g}(\theta, \mu)  \tilde{R}_{x_1,x_2}^{y_1,y_2}(c(\theta, \mu), \nu(\theta, \mu))    \, .
%	\quad x_1, x_2, y_1, y_2 \in \mathbb{R} \  ,
%\end{equation}
%that becomes   a $S$-matrix  if we replace  in \eqref{m7},  $g(z)$ by $\tilde{g}(z)$ and
%$\cot^2(z)$ by $1/c^2(z, \mu)$.
%the following conditions
 %
 %\barray
 %\tilde{g}^*(\theta, \mu) & =  &  \tilde{g}( - \theta^*, \mu) , \quad \lim_{\theta \rightarrow 0}  \tilde{g}(\theta, \mu)  = 1,
 %\quad \tilde{g}(\theta, \mu) \tilde{g}(-\theta, \mu)  = 1 \ ,
 %\label{s58}
 %\earray
 %
% and
 %
 % \beq
 % \tilde{g}(i \pi - \theta, \mu)  =    \frac{1}{ c^2(\theta, \mu)}  \tilde{g}(\theta, \mu)  \ .
 % \label{s59}
 %\eeq

\begin{figure}[t!]
\vspace{.2cm}
\begin{center}
\includegraphics[width=0.4\textwidth]{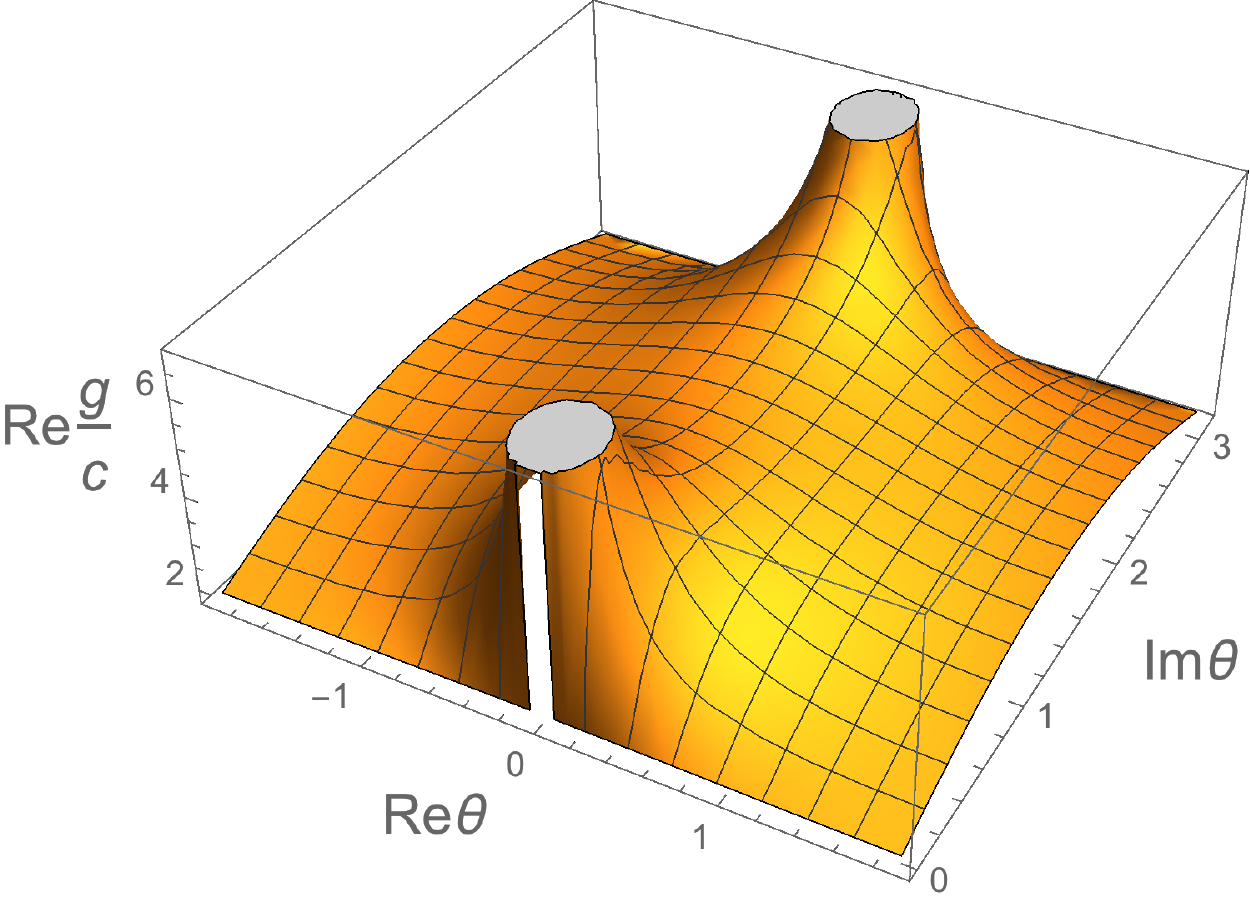} \qquad
\includegraphics[width=0.4\textwidth]{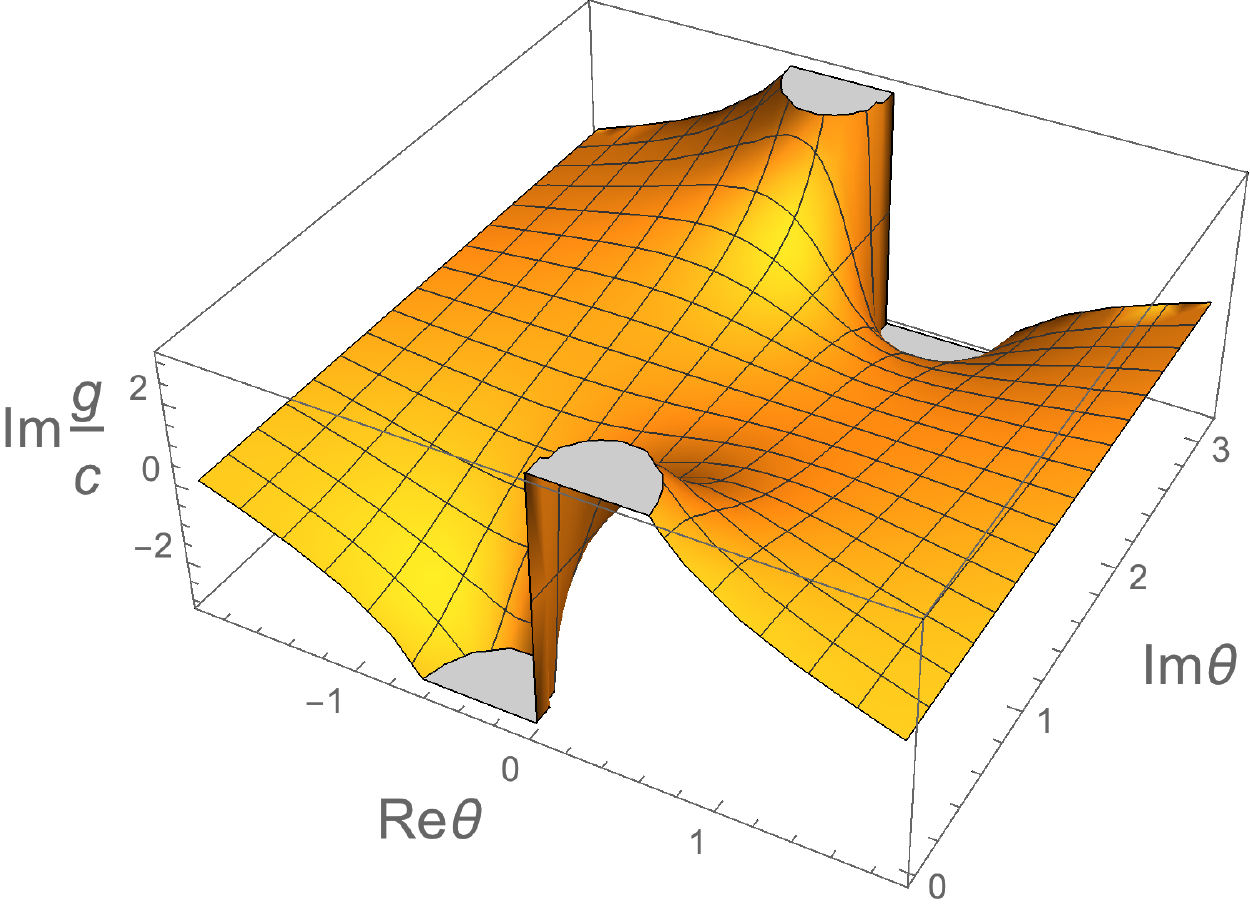}
\includegraphics[width=0.4\textwidth]{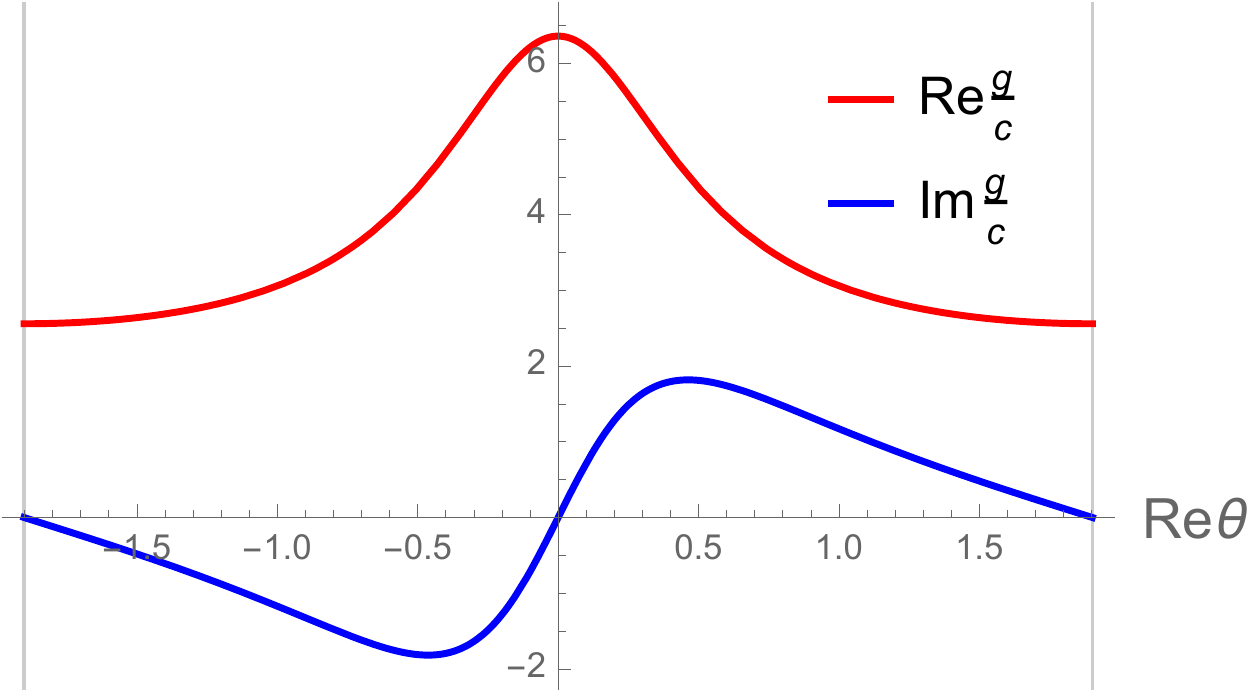}
\caption{
Above: 3-dimensional plot of the real and imaginary parts of $g(\theta,\mu)$ along a periodic cycle, $|{\rm Re}(\theta) |\leq {\tau \over 2}$, for $\mu=0.3$. Below: Section of the previous plots at ${\rm Im}(\theta)={\pi \over 6}$.
  }
\label{gM}
\end{center}
\end{figure}

The knowledge of the zeros and  poles of the Jacobi elliptic functions allows us to rewrite
%$c(\theta, \mu)$ as
 %
  %
 \beq
 c(\theta, \mu) = \lim_{M \rightarrow \infty}  \prod_{m = - M}^M \tan \left( \frac{ \theta+ m \tau}{2 i} \right)
% \frac{ \left(  n + i m \frac{K'}{2K} - {\theta \over 2 \pi i} \right)  } { \left(  n + \frac{1}{2} + i m \frac{K'}{2K} + {\theta \over 2 \pi i} \right)} \ .
%\frac{  n  - {1 \over 2 \pi i} (\theta\!+\!m \tau)   } {  n + \frac{1}{2}  + {1 \over 2 \pi i}(\theta\!-\!m \tau) } \ .
 \label{s60}
 \eeq
%
%  \beq
% c(\theta, \mu) = \prod_{n, m = - \infty}^\infty
% \frac{ \left(  n + i m \frac{K'}{2K} - {\theta \over 2 \pi i} \right)  } { \left(  n + \frac{1}{2} + i m \frac{K'}{2K} + {\theta \over 2 \pi i} \right)} \ .
%\frac{  n  - {1 \over 2 \pi i} (\theta\!+\!m \tau)   } {  n + \frac{1}{2}  + {1 \over 2 \pi i}(\theta\!-\!m \tau) } \ .
% \label{s60}
% \eeq
  %
with $\tau\!=\!{\pi K(\mu_1) \over K(\mu)}$.
Using this equation one finds the solution (see Appendix D)
 \beq
g(\theta, \mu) =    \lim_{M \rightarrow \infty} \!\!
 \prod_{n, m = - M}^M    \left[
% \frac{ \Gamma \left( n + i m \frac{ K'}{2K} + {\theta \over 2 \pi i} \right) \Gamma \left( n + \frac{1}{2} + i m \frac{ K'}{2K}  - {\theta \over 2 \pi i} \right) } {  \Gamma  \left( n + i m \frac{ K'}{2K} -  {\theta \over 2 \pi i} \right)  \Gamma \left( n + \frac{1}{2} + i m \frac{ K'}{2K}  + {\theta \over 2 \pi i} \right)}  \right]^2  \!  .
\frac{ \Gamma \left( n  + {1 \over 2 \pi i}(\theta\!-\!m \tau)  \right) \Gamma \left( n + \frac{1}{2} - {1 \over 2 \pi i} (\theta\!+\!m \tau) \right) } {  \Gamma
\left( n  - {1 \over 2 \pi i} (\theta\!+\!m \tau)  \right)  \Gamma \left( n + \frac{1}{2} + {1 \over 2 \pi i}(\theta\!-\!m \tau)  \right)}  \right]^2  \!  .
 \label{s61}
  \eeq
In the $\mu \to 0$ limit $\tau$ diverges and trivializes all contributions $m \neq 0$, reducing this expression to \eqref{m20}.  Compared to it, $g(\theta,\mu)$ is also meromorphic with additional poles and zeros, but all of them lie  outside the physical strip or at its boundary.  Applying a discrete version of \eqref{m21},  equation \eqref{s61} can be expressed as
  \barray
% \tilde{g}(z, \mu) &  =  & {\rm exp} \left[  \frac{ 2 \pi K z}{ K'} +  \sum_{n= 1 }^\infty   \frac{1}{ n} \frac{ \sinh( 4 \pi K z n/K')}{ \cosh^2( \pi K n/K')} \right]  \ .
%g(\theta, \mu) &  =  & {\rm exp} \left[ - i \left( \frac{ K \theta}{ K'} + \sum_{n= 1 }^\infty   \frac{1}{ n} \frac{ \sin( 2 \theta K n  /K')}{ \cosh^2( \pi K n/K')} \right) \right]  \ .
g(\theta, \mu) &  =  & {\rm exp} \left[ - i \left( \frac{ \pi \theta}{ \tau} + \sum_{n= 1 }^\infty   \frac{1}{ n} \frac{ \sin( 2 \pi \theta n  /\tau)}{ \cosh^2( \pi^2 n/\tau)} \right) \right]  \ .
 \label{s62}
  \earray
%which is a discrete version of \eqref{m21}.
The main difference between the $\mu=0$ and $\mu \neq 0$ S-matrices
%massive and massless cases
is the cyclic structure, with period $\tau$, introduced on the physical strip by the elliptic deformation. In Fig.\ref{gM} we have plotted the combination ${g\over c}$ in the region $|{\rm Re}(\theta) |\leq {\tau \over 2}$. Its structure indeed replicates that of Fig.\ref{g}.
%\textcolor{red}{more}.

 In connection with  the $S$-matrix described above,
 we would like to mention  another  models  whose $S$-matrices are expressed in terms of Jacobi elliptic
 functions. Zamolodchikov constructed
 a $S$-matrix  with a $\mathbb{Z}_4$ symmetry,  which is doubly periodic in the rapidity and depends on two coupling constants, being
 one of them the modulus  of the elliptic functions \cite{Z79}.
 In the limit where the modulus  vanishes
 one recovers  the  $S$-matrix of the sine-Gordon model with $O(2)$ symmetry.
 The model possess an infinite number of resonances related to the elliptic modulus.
 At high energies ($s  \gg m^2$),  the amplitudes and cross-sections are periodic in
 $\log s$,  which is a characteristic feature of renormalizable quantum field theories  with
 limit cycles \cite{W71}.  This is the field theory interpretation proposed in \cite{Z79} to correspond
 to the $S$-matrix. This reference also notices  the formal relation between the
 elliptic $S$-matrix and the Baxter's  eight vertex model  \cite{B72}.
% implying  that the infinite volumen  partition function of the latter model, should
% arise in solving the unitarity and analiticity of the corresponding $S$-matrix.
% }

Another example of elliptic $S$-matrix was  proposed by  Mussardo and Penati
that contains only one type of fundamental  particle with $S$-matrix and $\mathbb{Z}_2$ symmetry  \cite{M00}
\beq
S(\theta) =   \frac{ {\rm sn} \left( \frac{ K (\theta - i \pi a)}{i \pi}  \right)
{\rm cn} \left( \frac{ K (\theta + i \pi a)}{i \pi}  \right) {\rm dn} \left( \frac{ K (\theta + i \pi a)}{i \pi}  \right)}{
 {\rm sn} \left( \frac{ K (\theta + i \pi a)}{i \pi}  \right)
{\rm cn} \left( \frac{ K (\theta - i \pi a)}{i \pi}  \right) {\rm dn} \left( \frac{ K (\theta - i \pi a)}{i \pi}  \right)}
\label{MP}
\eeq
where $a$ is a coupling constant in the interval $[0,1/2]$.
%Equation \eqref{MP} exhibits the duality $a \leftrightarrow 1-a$.
This $S$ is periodic in $\theta$ also with period $\tau$
and correspondingly  an infinite number of unstable resonances.
In the limit where the elliptic modulus goes to zero, one recovers the $S$-matrix of the sinh-Gordon model.
In reference \cite{M00}, it is conjectured  that the $S$-matrix \eqref{MP}  corresponds in the UV
to a non unitary and irrational CFT. Notice that \eqref{MP} can be written
as $S(\theta) = \frac{c(\theta- i \pi a)}{c(\theta + i \pi a)}$.
% If $a=0$, obviously $S(\theta) =1$,
%For $a=0$, \eqref{MP} contains,  in the numerator and denominator,
%the factor $c(\theta, \mu)$,  given in eq.\eqref{cu}, such that  $S(\theta) =1$.
%This is in contrast with  our $S$-matrix that depends on the factor $g/c$, that makes the scattering non-trivial.
This  relation suggests a possible  deformation of our model with a  parameter similar to  $a$ in \eqref{MP}.

Finally, there  are another  models with continuous variables living on a circle
whose Boltzmann weights satisfy the star triangle relation, and are
expressed in terms of the elliptic gamma-function  \cite{BS12a,BS12b}.

\subsection{The $S$-matrix  in Fourier space}

The two particle scattering equation \eqref{s0} can be given an algebraic form due to  Faddeev and Zamolodchikov
in terms of operators ${\bf A}_i(\theta)$  with $\theta$ real,  whose  action  on the Hilbert space
vacuum $|0 \rangle$ creates a particle with rapidity $\theta$,
 %in the state $| A_i(\theta) \rangle$,
%
\beq
{\bf A}_i(\theta) |0 \rangle =
| A_i(\theta) \rangle   \, .
\label{s0a}
\eeq
The bosonic model has associated a continuous set of such operators ${\bf A}_x(\theta)$.
Equation \eqref{s0} is  equivalent to the exchange  relation
\beq
{\bf A}_{x_1} ( \theta_1) {\bf A}_{x_2} ( \theta_2) =    \int d y_1 d y_2  \; {S}_{x_1  x_2}^{y_1 y_2}(\theta_{12})  \,
{\bf A}_{y_1} ( \theta_2) {\bf A}_{y_2} (\theta_1)  \, .
\label{f3}
\eeq
Explicit realizations of field theory operators satisfying this relation are generally unknown.
We next  propose a  partial  realization of them,
which in the massless limit corresponds to the vertex operators of the CFT describing a massless boson.

Let us first define the Fourier transform of the  Faddeev-Zamolodchikov   operators %${\bf A}_x(\theta)$
\beq
{\bf \hat{A}}_q( \theta) = \int  d x \,  e^{ i q x} {\bf A}_x(\theta), \qquad q \in \mathbb{R} \ ,
\label{f1}
\eeq
where the integral runs over the real line.  Recalling that $x$ represents the scalar field $\phi$,
we interpret  $q$ as a charge associated to the symmetry $\phi \rightarrow \phi + {\rm cte}$
of the massless $S$-matrix. We shall find below further support of  this interpretation.
The operators ${\bf \hat{A}}_q( \theta)$ satisfy the following exchange relation derived from \eqref{f3}
\barray
{\bf \hat{A}}_{q_1} ( \theta_1) {\bf \hat{A}}_{q_2} ( \theta_2) &  =  &
  \int d p_1 d p_2  \,  \hat{S}_{q_1  q_2}^{p_1 p_2}(\theta_{12})
{\bf \hat{A}}_{p_1} ( \theta_2) {\bf \hat{A}}_{p_2} (\theta_1) \ ,   \label{f4}
\earray
%with
where $\hat{S}_{q_1  q_2}^{p_1 p_2}$ is the Fourier transform of $S_{x_1 x_2}^{y_1 y_2}$. Diagonalizing  the quadratic form in the exponent of the R-matrix
\eqref{Rm} before Fourier transforming, we easily obtain (see Appendix D)
\barray
\hspace{-1.8cm} \hat{S}_{q_1  q_2}^{p_1 p_2} &  \hspace{-.9cm}= &\!\!\!\!{g \mu_1 \over 8 \pi c \sqrt{\mu}} \; {\rm exp} \! \left[  -\frac{1}{4} \! \left( \frac{1}{c {\widetilde m}^2} \,( q_1 \!+\! q_2\! -\! p_1\! -\! p_2)^2 + \frac{c}{4 \!+ \!c^2 {\widetilde m}^2} \,( q_1 \!+\! q_2\! +\! p_1\! +\! p_2)^2  \right. \right.  \\
&& \hspace{1.1cm} \left. \left. + \; \frac{1}{ c (4\! +\! {\widetilde m}^2)}\, (q_1 \!-\! q_2\! -\! p_1\! +\! p_2)^2 + \frac{ c}{ 4\! +\! c^2 (4\!+ \!{\widetilde m}^2)} \, (q_1\! -\! q_2 \!+\! p_1 \!-\! p_2)^2 \right) \right] \ .
\nonumber
\earray
In the trigonometric limit $\widetilde m$ vanishes and the first term in the exponential gives rise to a delta function
 \barray
\label{i11}
 \hspace{-1.7cm}  \hat{S}_{q_1 q_2}^{p_1 p_2} & \hspace{-.8cm} =  &  \hspace{-.4cm} \delta(q_1 \!+\! q_2 \!-\! p_1 \!-\! p_2) \;  i g \coth{\theta \over 2}
   \left( {\sinh \theta \over 8 \pi i } \right)^{\!1/2} \\[1mm]
 & \!\!\!\!  \times &  {\rm exp} \left[  \frac{i}{4}  \left(  \tanh \frac{\theta}{2}   (q_1\! +\! q_2)^2 - \coth \frac{\theta}{2}   (q_1\! -\! p_1)^2 +  \frac{1}{2} \sinh \theta   (q_1\! -\! p_2)^2 \right) \right]    \ .
 \nonumber
 \earray
It implies that the scattering preserves the total  charge, that is
 $q_1 + q_2 = p_1+p_2$. This
 property is due to the invariance of the trigonometric $S$-matrix  under the
 shift of all the variables.
 Moreover,
in the limit of large rapidity  $\theta \rightarrow  \pm \infty$ a new delta function emerges.
The combination $i g \coth {\theta\over 2}$ at the same time  tends to $1$, as seen in Fig.\ref{g}, obtaining
\begin{equation}
\label{i12}
\lim_{\theta \rightarrow  \pm \infty }
 \hat{S}_{q_1 q_2}^{p_1 p_2}(\theta)  =      \delta(q_1  - p_2)  \; \delta(q_2  - p_1) \, e^{ \pm i q_1 q_2}  \ .
 \end{equation}
The exchange equation \eqref{f4} reduces then to
\begin{equation}
{\bf \hat{A}}_{q_1} ( \theta_1) {\bf \hat{A}}_{q_2} ( \theta_2)   \rightarrow
 e^{ \pm i q_1 q_2}  {\bf \hat{A}}_{q_2} ( \theta_2) {\bf \hat{A}}_{q_1} ( \theta_1)  \ , \qquad \theta_{12} \to \pm \infty  \ .
 \label{i13}
 \end{equation}
 This expression is similar to  the braiding of chiral  and antichiral
 vertex operators $e^{i q \phi(z)}$ and  $e^{i q \bar{\phi}(\bar{z})}$ in the $c=1$ CFT of a massless boson \cite{CFT}.
 The result above suggests the existence of an explicit form of
${\bf \hat{A}}_q(\theta)$ interpolating between the chiral and antichiral vertex operators for generic rapidity.
% This   behaviour is  proper of a massless theory in the far UV is a consistency check for the scattering interpretation.
%\textcolor{blue}{The emergence of behaviour proper of a massless theory in the far UV suggests that
%the field theory underlying  the trigonometric  model is given by a  massless boson  perturbed by a relevant
%operator that drives the model to a massive regime. This interpretation could be supported
%by a TBA calculation but this task goes beyond the scope of this work.}
% The result above suggests the existence of an explicit form of
%${\bf \hat{A}}_q(\theta)$ interpolating between the chiral and antichiral vertex operators for generic rapidity.
%This also would imply that at intermedium and low scales the theory is not critical, which should trace back to the lattice structure of
%the original Stat. Mech. model. In order to better understand the theory in the IR we would need to construct its associated TBA.
%We will not address however this question here.

In the elliptic case the rapidity becomes cyclic with period $\tau$.
%The boundaries of a cycle, $\theta=\pm {\tau \over 2}$, represent the UV limit of this theory. There the S-matrix is
% On the other hand, in the massive case $\theta \in [- \frac{\tau}{2}, \frac{\tau}{2}]$.
At the boundaries of the cycle, $\theta=\pm {\tau \over 2}$, we have
 %
 %\beq
 %h( \pm \tau/2  ) = \sqrt{  \frac{ 1 - \sqrt{\mu}}{ 1 + \sqrt{\mu}}  }, \qquad  \bar{\nu}( \pm \tau/2 )  = 4 \sqrt{   \frac{ \mu}{ \mu_1}} \ ,
%c( \pm \tau/2 ,\mu ) = \mp i \frac{ \sqrt{\mu_1}}{ 1 + \sqrt{\mu}}  , \qquad  \nu( \pm \tau/2 )  = \mp 4i \sqrt{   \frac{ \mu}{ \mu_1}} \ ,   g=\mp i
%  \label{i14}
%  \eeq
% and hence
 %
% \barray
%\label{i12}
%\hat{S}_{q_1  q_2}^{p_1 p_2}\Big(\!\! \pm {\tau \over 2}\Big) &   =(1\!+\!\sqrt{\mu}) {\sqrt{\mu_1} \over 8 \pi \sqrt{\mu}} \,
%\propto
%e^{  \pm i \frac{ \sqrt{\mu_1}}{4} \left[ (q_1 + p_2) (q_2 + p_1) - \frac{1}{\sqrt{\mu}}   (q_1 - p_2) (q_2 - p_1) \right] }  \ .
% \earray
\begin{equation}
\label{i14}
\hspace{15mm} \hat{S}_{q_1  q_2}^{p_1 p_2}\Big(\!\! \pm\! {\tau \over 2}\Big)   = (1\!+\!\sqrt{\mu}) \, e^{  \pm i \frac{ \sqrt{\mu_1}}{4} (q_1 + p_2) (q_2 + p_1) }
\times \left( {\sqrt{\mu_1} \over 8 \pi \sqrt{\mu}} \,e^{  \mp i \frac{ \sqrt{\mu_1}}{4 \sqrt{\mu}}  (q_1 - p_2) (q_2 - p_1)  } \right) \ .
 \end{equation}
 %In this case there is not a simple interpretation in terms of vertex operators, which is consistent with the fact that this $S$ matrix should  describe a massive theory.
 Writing the exponent of last term in parenthesis as the difference of two squares, we observe that it defines
%The last term in parenthesis is
a gaussian distribution centered around $q_1=p_2$ and $q_2=p_1$ with broadness ${8 \sqrt{\mu} \over \sqrt{\mu_1}}$.
%the exponent in \eqref{i14} can be written as the difference of two squares and then one gets the \eqref{i12}.  }
%There is no a simple interpretation in terms of vertex operators for \eqref{i14}.
This is consistent
% with the fact that this $S$ matrices should  describe a massive theory.
with the periodic structure in the real direction of the rapidity,
which acts as an effective UV cutoff and implies that there is no limit in the theory were the mass completely decouples.
In the limit $\mu \to 0$ the gaussian distribution tends to a product of delta functions, recovering \eqref{i12}.

\section{Conclusions}

In this paper we have applied the theory of exactly solvable  models to a massless and massive boson living
in a two dimensional square lattice. We have shown that the Boltzmann weights satisfy the Yang-Baxter equation, with the difference
property in the rapidity variable,
using a parameterization in terms of trigonometric functions in the massless model,  and Jacobi elliptic functions in the massive model.
In the former case,  the partition function is invariant under the shift of the scalar field $\phi \rightarrow \phi + {\rm constant}$,
while in the latter case it has the $\mathbb{Z}_2$ invariance $\phi \rightarrow - \phi$.  These properties are reminiscent
of the Boltzmann weights of the 6 vertex and 8 vertex models. We have  calculated the eigenvalues and eigenvectors
 of the row-to-row transfer matrix,
and the corresponding conserved quantities, that were also obtained  using the Quantum Inverse Scattering method.
In the massless case the connection with the $c=1$ CFT  was established.
Finally, starting from the Boltzmann weights of the Statistical Mechanics model, we have proposed a scattering theory
for massive particles with  a continuous degree of freedom that satisfy  all the  standard axioms.
We conjecture
that the trigonometric $S$-matrix corresponds in the UV to a relevant perturbation of the $c=1$ massless CFT.
The field theory associated to the elliptic solution is more difficult to interpret due to the presence of a UV cutoff
related to  the modulus of the elliptic solutions. A possibility,  along the lines of the Zamolodchikov
elliptic $S$-matrix model with  $\mathbb{Z}_4$ symmetry,  is that  this field theory has limit cycles.
%Further investigation is required to clarify this problem.

In this work we have considered a discretized free field theory,
but the method can be in principle applied to lattice versions of integrable field theories.  From another viewpoint,
 the  results presented in this work could have applications to encode quantum information in continuous degrees of freedom \cite{G01},
and the design of quantum circuits  that enjoy an analog of crossing  symmetry.
Further investigations  are  required  to clarify all these issues.

%(\textcolor{red}{A�adir alguna referencia de QC sobre este tema}).

\section*{Acknowledgements}

We would like to thank Francisco  Alcaraz, Giuseppe Mussardo, Adri\'an Franco-Rubio,
Paul Pierce and Guifr\'e Vidal for discussions. This work has been financed by the grants  PGC2018-095862-B-C21,
QUITEMAD+ S2013/ICE-2801, SEV-2016-0597 of the
``Centro de Excelencia Severo Ochoa'' Programme and the
CSIC Research Platform on Quantum Technologies PTI-001.

\section{Appendix A}

%\subsection{Yang–Baxter equation: massless case}

The massless Yang–Baxter equation can be constructed from \eqref{R} and \eqref{YB2}. The explicit form of this equation is given by the matrix defined in \eqref{YB3}. Its off-diagonal blocks are
\begin{gather}
	\label{explicitM1}
	M_{x y} =
		\begin{pmatrix}
			0 & 0 & 0 \\
			0 & 0 & 0 \\
			- {1 \over c_2} & 0 & 0
		\end{pmatrix}
	\ , \
	M_{x z} =
		\begin{pmatrix}
			0 & - c_1^{-1}  & 0 \\
			- c_1^{-1}  & 0 & 0 \\
			- c_2 & 0 & 0
		\end{pmatrix}
	\ , \
	M_{y z} =
		\begin{pmatrix}
			0 & 0 & - c_2 \\
			0 & 0 & - c_3^{-1} \\
			0 & - c_3^{-1} & 0
		\end{pmatrix}
	\ ,
\end{gather}
and $M_{i j} = M_{j i}^T$. The diagonal blocks are
\begin{gather}
	\label{explicitM2}
	M_{x x} =
		\begin{pmatrix}
			c_1 + c_1^{-1} & - c_1 & 0 \\
			- c_1 & c_1 + c_1^{-1} & 0 \\
			0 & 0 & c_2 + c_2^{-1}
		\end{pmatrix}
	\ , \
	M_{y y} =
		\begin{pmatrix}
			c_2 + c_2^{-1} & 0 & 0 \\
			0 & c_3 + c_3^{-1} & - c_3 \\
			0 & - c_3 & c_3 + c_3^{-1}
		\end{pmatrix}
	\ , \\
	M_{z z} = M_{x x} + M_{y y} + M_{x y} + M_{x y}^T
	\ .
\end{gather}

When the integration over the $z$ variables in \eqref{YB3} is done, we obtain the matrix $N$ defined in \eqref{YB33}. Its blocks are given by
\begin{gather}
	\label{Nmatrix1}
		N_{x x}
	=
		M_{x x} - M_{x z} M_{z z}^{-1} M_{z x}
	\ , \\
		N_{y y}
	=
		M_{y y} - M_{y z} M_{z z}^{-1} M_{z y}
	\ , \\
	\label{Nmatrix3}
		N_{x y}
	=
		N_{y x}^T
	=
		M_{x y} - M_{x z} M_{z z}^{-1} M_{z y}
  \ .
\end{gather}
The massless Yang-Baxter equation is then reduced to conditions $N_{xx}=N_{yy}$ and $N_{xy}=N_{yx}$. This system of equations has a unique solution which is given by
\beq
	c_2 =  \frac{c_1 + c_3}{1- c_1 c_3}  \ .
\eeq

%\subsection{Yang–Baxter equation: massive case}

In the massive case, the analogues of \eqref{explicitM2} are given by
\begin{gather}
	\widetilde M_{x x} = M_{x x} +
		{1 \over 2} \begin{pmatrix}
			c_1 \widetilde m_1^2 & 0 & 0 \\
			0 & c_1 \widetilde m_1^2 & 0 \\
		 0  & 0 & c_2 \widetilde m_2^2
		\end{pmatrix} \ , \;\;
%	\\
	\widetilde M_{y y} = M_{y y} +
		{1 \over 2} \begin{pmatrix}
			c_2 \widetilde m_2^2 & 0 & 0 \\
			0 & c_3 \widetilde m_3^2 & 0 \\
			0 & 0 & c_3 \widetilde m_3^2
		\end{pmatrix}
	\ , \\
	\widetilde M_{z z} = M_{z z} +
		{1 \over 2} \begin{pmatrix}
			c_1 \widetilde m_1^2 + c_2 \widetilde m_2^2 & 0 & 0 \\
			0 & c_1 \widetilde m_1^2 + c_3 \widetilde m_3^2 & 0 \\
			0 & 0 & c_2 \widetilde m_2^2 + c_3 \widetilde m_3^2
		\end{pmatrix}
	\ .
\end{gather}
The other blocks remain identical $\widetilde M_{i j} = M_{i j}$, $i \neq j$. After the integration of the $z$ variables is performed, the resulting blocks are decribed by an equation analogous to \eqref{Nmatrix1}-\eqref{Nmatrix3} but replacing $M_{i j} \to \widetilde M_{i j}$. In this case, conditions $N_{xx}=N_{yy}$ and $N_{xy}=N_{yx}$ form a system of equations which has the solution given by \eqref{l1}-\eqref{l3}.

\section{Appendix B}

%\subsection{Transfer matrix}

The expectation values of the coordinate transfer matrix are codified in \eqref{T} in terms of a matrix $M$ analogous to the ones used in the Appendix A.
Its blocks are
\begin{gather}
		M_{x x} = M_{y y} = {1 \over 2 } M_{zz} = a \, \id
	\ , \quad   a= {1 \over c} + c + { c \, \widetilde m \over 2}
	\ , \\
		M_{x y}=0 \ , \quad
		M_{x z} = - {1 \over c} S - c \id
	\ , \quad
		M_{y z} =  - {1 \over c} \id - c S
	\ ,
\end{gather}
where the shift matrix is defined as $S_{ij}=\delta_{i,j+1}$, with $L+1 \equiv 1$. When the integration of the $z$ variables is done, we obtain a matrix $N$ with the same structure as \eqref{Nmatrix1}-\eqref{Nmatrix3}. The computation of this matrix is straightforward, its explicit form is shown in \eqref{T21}-\eqref{T22}.

%\subsection{Expansion of a gaussian in delta functions}

The elements of the transfer matrix can be written as the product of two factors, $T_p$ and $T_q$, defined in \eqref{Tpp}-\eqref{Tqq}. When the lattice anisotropy $c$ is small, $T_p$ becomes a sharply picked gaussian which can be expanded in terms of delta functions.
%Our goal in this appendix is to expand a gaussian function inside an integral by a series of derivatives of delta functions. Firstly,
Let $f(x)$ be an analytic function, then we can expand
%any smooth function $f(x)$ can be substituted by its polynomial expansion in the following expression
\begin{equation}
	\label{fexpansion}
		{1 \over \sqrt{\pi c }} \int \dd x\,  e^{ - {x^2 \over c} } f(x) =
%{\p_x^{2k} f(0)	=
		{1 \over \sqrt{\pi c}} \sum_{k=0}^\infty  {f^{2k)}(0) \over (2k)!} \int \dd x \, x^{2k} e^{ - {x^2 \over c} }
	\ .
\end{equation}
The integrals can be easily performed, and the derivatives of the function f reexpressed as follows
\begin{equation}
 f^{2k)}(0) = \int \! \dd x \, \delta^{2k)} (x) \, f(x)  \ .
\end{equation}
Substituting this in \eqref{fexpansion} we obtain
%The derivatives of $f(0)$ can be substituted by delta functions $ \p_x^{2k} f(0) = \int \! d x \, \delta^{2k)} (x) \, f(x) $ and the integral performed to get
\begin{equation}
		{1 \over \sqrt{\pi c}} \int d x \ e^{ - {x^2 \over c} } \ f(x)
	=
%		\sum_{k=0}^\infty {(2k-1)!! \ c^k \over (2k)! \ 2^k} \int \dd x \ \delta^{2k)}(x) \ f(x)
		\sum_{k=0}^\infty { c^k \over k! \ 4^k} \int \dd x \ \delta^{2k)}(x) \ f(x) \ ,
\end{equation}
which in the limit $c \to 0$ implies the desired expansion
\begin{equation}
		{1 \over \sqrt{\pi c}} e^{ - {x^2 \over c} }
	=
		\delta(x) + {c \over 4} \delta''(x) + \order{c^2}
	\ .
\end{equation}

%\subsection{Eigenstate condition}

Finally, we will clarify some technical details on the diagonalization of the transfer matrix performed in Section  \ref{coordinateTM}.
Using the ansatz
\begin{equation}
\ket{ \Psi } = \int \dd {\vec x} \, f_n ({\vec x}) \,e^{ - {1 \over 2} {\vec x} \, K \, {\vec x}^T } \, \ket{ \vec x } \ ,
\end{equation}
and the transfer matrix elements \eqref{T2}-\eqref{T22}, the eigenstate condition ${\bf T}(u) \ket{ \Psi } = \Lambda \ket{ \Psi }$ becomes
%implies \eqref{hermite} and \eqref{eigenmatrix}. The definition of ${\bf T}(u)$ is given in \eqref{T2} and the definition of $\ket{ \Psi }$ in \eqref{eigen}. Let us repeat them here
%\begin{align}
%		\bra{ \vec y } {\bf T} (u) \ket{ \vec x }
%	=
%		{ 1 \over (4 \pi a c^2)^{L\over 2}}    \,
%		e^{ - {1 \over 2}( \vec{ x},\vec{ y} )  \,  N \, ( \vec{ x},\vec{ y}  )^T }
%	\ , \\
%		\ket{ \Psi }
%	=
%		\int \dd {\vec x} \,
%			f_n ({\vec x}) \,e^{ - {1 \over 2} {\vec x} \, K \, {\vec x}^T } \, \ket{ \vec x }
%			\ .
%\end{align}
%The explicit form of \eqref{eigencond} in terms of the fields is
\begin{equation}
	\label{eigencond2}
		{ 1 \over (4 \pi a c^2)^{L\over 2}}
		\int \dd {\vec x} \,
		f_n ({\vec x}) \,
		e^{ - {1 \over 2} \vec x ( K + N_1 ) \vec x^T - {1 \over 2} \vec y N_1 \vec y^T - \vec{ x } N_2 \vec{ y }^T }
	=
		\Lambda
		f_n ({\vec y}) \,
		e^{ - {1 \over 2} {\vec y} \, K \, {\vec y}^T }
	\ .
\end{equation}
%where we have used the definitions $N_1 = N_{x x} = N_{y y}$ and $N_2 = N_{x y} = N_{y x}^T$ from \eqref{T21}-\eqref{T22}.
In order to perform the integration over the $x$ variables, %in the \emph{lhs} of \eqref{eigencond2},
we make the shift $\vec x \to {\vec x}-{\vec y} \,N_2^T (N_1 + K)^{-1}$ and obtain
\begin{equation}
	\label{eigencond3}
		{ 1 \over (4 \pi a c^2)^{L\over 2}}
		\int \dd {\vec x} \,
		f_n \big({\vec x}-{\vec y} \,N_2^T (N_1 + K)^{-1} \big)
		e^{ - {1 \over 2} \vec x ( K + N_1 ) \vec x^T }
	=
		\Lambda
		f_n ({\vec y}) \,
		e^{ - {1 \over 2} {\vec y} \, \widetilde K \, {\vec y}^T }
	\ ,
\end{equation}
where $\widetilde K = K - N_1 + N_2^T (N_1 + K)^{-1} N_2$. This relation implies the following two conditions
%From \eqref{eigencond3} we read the two conditions
\begin{align}
K - N_1 + N_2^T (N_1 + K)^{-1} N_2 & = 0  \ ,\\
		\int \dd {\vec x} \,
		f_n \big({\vec x}-{\vec y} \,N_2^T (N_1 + K)^{-1} \big)
		e^{ - {1 \over 2} \vec x ( K + N_1 ) \vec x^T }
	& =
		(4 \pi a c^2)^{L\over 2} \, \Lambda \,
		f_n ({\vec y}) \, \label{eigen4}
	%\\
%	K - N_1 + N_2^T (N_1 + K)^{-1} N_2 & = 0
	\ .
\end{align}
The first equation leads to the solution \eqref{Q} for $K$. The second allows to obtain both the functions $f_n$ and the eigenvalues $\Lambda$.
We make an ansatz for $f_n$ based on the Hermite polynomials
%
%\subsection{Excited states}
%
%In this appendix we are going to use the ansatz \eqref{fnhermite} in \eqref{eigencond4} to obtain an explicit formula for $\Lambda$. Let us recall here that the ansatz are the Hermite polynomials
\begin{equation}
	\label{herans}
	f_n({\vec x}\, ;{\vec v}) = \rho \, e^{  \, {\vec x} \, K \, {\vec x}^T } \left( \! {\vec v}.{\partial \over \partial {\vec x}}  \right)^{\!\!n} e^{ -  {\vec x} \, K \, {\vec x}^T }  \ ,
\end{equation}
with $\rho$ a normalization constant and ${\vec v}$ a vector to be determined. This expression is inserted in \eqref{eigen4} and the integration variables shifted $\vec x \to \vec x - 2 \vec y N_2^{-1} K$ to rewrite the \emph{lhs} as
\begin{equation}
	\rho \,
	e^{ \vec y K \vec y^T }
	\int \dd {\vec x} \,
	e^{ - {1 \over 2} \vec x ( K - N_1 ) \vec x^T }
	\left( \!
		{\vec v}.{\partial \over \partial {\vec x}}
	\right)^{\!\!n}
	e^{ -  {\vec z} \, K \, {\vec z}^T }
	\ ,
\end{equation}
where $\vec z = \vec x - \vec y N_2^{-1} (N_1 + K)$. Using now that $\partial_{\vec x} f(\vec z) = - (N_1 + K)^{-1} N_2 \cdot \partial_{\vec y} f(\vec z)$,
the derivative can be brought outside the integration. We can then easily perform the gaussian integration, reducing \eqref{eigen4} to
\begin{equation}
		\Lambda_0 \,
		\rho \,
		e^{ \vec y K \vec y^T }
		\left( \!
			- {\vec v} (N_1 + K)^{-1} N_2^T . {\partial \over \partial {\vec y}}
		\right)^{\!\!n}
		e^{ - \vec y K \vec y^T }
	=
		\Lambda \,
		f_n({\vec y}\, ; \vec v)
	\ .
\end{equation}
The constant $\Lambda_0$ is the eigenvalue of the transfer matrix on the ground state, $f_0=1$, given in \eqref{L0}.
This equation is satisfied provided $\vec v$ is an eigenstate of the matrix $ -(N_1 + K)^{-1} N_2$, whose eigenvalues are
\begin{equation}
\lambda= e^{i p} {(1+c^2 e^{-i p})^2 \over c^2 (a+\omega)^2} \ ,
\end{equation}
with $p={2 \pi k \over L}$ and $k=0,..,L-1$. Then $\Lambda=\Lambda_0 \lambda^n$.
%This expression directly leads to \eqref{eigenvaluehermite}.

\section{Appendix C}

%\subsection{Conserved charges}

The model we are considering has an infinite tower of conserved charges derived from the expansion of the transfer matrix. We proposed in \eqref{Qeven} and \eqref{Qodd} the following operator form for the charges
%defined in \eqref{Qeven} and \eqref{Qodd}. The aim of this appendix is to check their commutation properties. Let us repeat the definitions:
\begin{equation}
	{\bf Q}_{2a} = {(-\!1)^{a} \over a } \, a_x \, {\vec {\boldsymbol \pi}}  \left( S^{aT} -S^{a}  \right)  {\vec {\bf x}}
	\ , \quad
	{\bf Q}_{2a+1} = {1 \over 2a\!+\!1} \Big( a_x^2 \, {\vec {\boldsymbol \pi}} \, K^{2a} \, {\vec {\boldsymbol \pi}} + {\vec {\bf x}} \, K^{2a+2} \, {\vec {\bf x}} \Big)
	\ .
	\label{charges2}
\end{equation}
The aim of this appendix is to check that they indeed commute among themselves, and that their expectation values agree with \eqref{vevQ}.

%Their commutators among all conserved charges must vanish.
The vanishing commutation between even charges can be easily derived from
\begin{equation}
\left[ \, {\vec {\boldsymbol \pi}} A \, {\vec {\bf x}} \,, {\vec {\boldsymbol \pi}} B \, {\vec {\bf x}}\, \right] =i {\vec {\boldsymbol \pi}} \, [A,B] \,{\vec {\bf x}} \ ,
\end{equation}
identifying $A=S^a\! -\!S^{a  T} $ and $B=S^b\! -\!S^{b  T}$ and using that $S$ and its transpose commute. Between odd charges from
\begin{equation}
\left[ \, {\vec {\bf x}} \, A \, {\vec {\bf x}} \, , {\vec {\boldsymbol \pi}} B \, {\vec {\boldsymbol \pi}} \, \right] =i {\vec {\boldsymbol \pi}} \, A (B+B^T)  \,{\vec {\boldsymbol \pi}}+ i{\vec {\bf x}} \, A (B+B^T)  \,{\vec {\bf x}} \ ,
\end{equation}
identifying $A=K^{2a+2}$ and $B=K^{2b}$ and using that $K$ is symmetric. Between even and odd charges from
\begin{equation}
\left[ \, {\vec {\bf x}} \, A \, {\vec {\bf x}} \, ,  {\vec {\boldsymbol \pi}} B \, {\vec {\bf x}}\, \right] = i{\vec {\bf x}} \, B (A+A^T) \,  {\vec {\bf x}}  \ , \qquad
\left[ \, {\vec {\boldsymbol \pi}} \, A \, {\vec {\boldsymbol \pi}} \, ,  {\vec {\boldsymbol \pi}} B \, {\vec {\bf x}}\, \right] = i{\vec {\boldsymbol \pi}}  (A+A^T) B \, {\vec {\boldsymbol\pi}}
\end{equation}
identifying $A=K^{2a}$ and $B=S^b\! -\!S^{b  T}$ and using that, since $K$ and $S$ commute, $(A+A^T) B=B(A+A^T) $ is an antisymmetric matrix.

The expectation values of the tower of conserved charges can be derived expanding the eigenvalues of the transfer matrix. In order to avoid powers of lower charges contributing to higher ones, we consider the logarithm of a generic eigenvalue
\begin{equation}
\log \Lambda = i p  -\sum_{k=1}^{L-1} (2n_k+1) \log c(a+\omega_k) +\sum_{k=1}^{L-1} 2n_k \log \big(1+c^2 e^{-i p_k}\big) \ ,
\end{equation}
The third term on the {\it rhs} only contains even powers of the uniformization parameter $u$, while the second contributes to both even and odd powers.
We can rewrite
\begin{equation}
\log \big(c(a+\omega_k)\big) = {1 \over 2} \left[ \log c^2(a^2-\omega_k^2) + \log {a+ \omega_k  \over a - \omega_k} \right] \ .
\end{equation}
Using \eqref{ac} and \eqref{omegak},
we can see that
\begin{equation}
c^2 (a^2 -\omega_k)^2 = (1+ c^2 e^{i p_k})(1+ c^2 e^{-i p_k}) \ .
\end{equation}
Substituting we have
\begin{equation}
\log \Lambda =i p \,-\,\sum_{k=0}^{L-1} \! \Big(n_k\!+\!{1 \over 2}\Big)  \log {a+\omega_k \over a- \omega_k} \,+\, \sum_{k=0}^{L-1} \!n_k \log {1+c^2 e^{-i p_k} \over 1+c^2  e^{i p_k}\;\;} -  \log \!\Big(1\! -\!(- c^2)^L \Big)\ .
\label{Lsimple2}
\end{equation}
The second term only contains now odd powers of $u$ and the third even ones. The second and third terms contribute therefore to different conserved charges. Using in $a^{-1}$ as expansion parameter for the second term  and $c^2$ for the third, we derive the following tower of conserved charges
\begin{equation}
%\langle {\bf Q}_{2l} \rangle= {i \over l } \sum_{k=1}^{L-1} n_k \sin( p_k l) \ , \hspace{.5cm} \langle {\bf Q}_{2l+1} \rangle= {1 \over 2 l\!+\!1 } \sum_{k=1}^{L-1} \Big(n_k\!+\!{1 \over 2}\Big) \,\omega_k^{2l+1}   \ ,
%\langle {\bf Q}_{2l} \rangle= {(-\!1)^l   \over l } \, \sum_{k=1}^{L-1} 2n_k \sin( p_k l) \ , \hspace{.4cm}
\langle {\bf Q}_{2a+1} \rangle= {2 \over 2 a\!+\!1 } \sum_{k=0}^{L-1} \Big(\!n_k\!+\!{1 \over 2}\Big)\,\omega_k^{2l+1}  \ ,\hspace{.4cm}  \langle {\bf Q}_{2a} \rangle= {(-\!1)^a   \over a } \, \sum_{k=1}^{L-1} 2n_k \sin( p_k a)  \ .
\label{vevQ2}
\end{equation}
Notice that $\langle {\bf Q}_{2a} \rangle$, with $a$ multiple of $L$, is trivially zero. There are however additional contributions to the eigenvalues of the transfer matrix at orders $c^{2Lb}$, coming from the last term in \eqref{Lsimple2}. These contributions are independent of the occupation numbers. They are also independent from momenta and energies, contrary to the vacuum piece of the odd charges vev. In terms of operators they must be associated with the identity, and hence are of no relevance to the integrability structure. For this reason we have dropped them in the main text, eq. \eqref{Lsimple}.

Finally we will show that the expectation values of the operators \eqref{charges2} agree with \eqref{vevQ2}. For simplicity, the check will be performed
only over 1-particle states. For them \eqref{herans} reduces to
\begin{equation}
f_1({\vec x}; {\vec v})= - 2 \rho \, {\vec v} K {\vec x} \ ,
\end{equation}
where $\vec v$ is an eigenvector $(N_1 + K)^{-1} N_2$, or equivalently, of the shift matrix $S$. Hence ${\vec v}={\vec v}_k$, with
$(v_k)_i={1\over \sqrt{L}} e^{i p_k}$. Using \eqref{cpi}, for the even charges we have
\begin{equation}
\langle {\bf Q}_{2a} \rangle_{1k} = -4 i \rho^2\!\! \int \dd {\vec x} \dd {\vec y} ({\vec v}_k K {\vec x})( {\vec v}_k^*  K {\vec y}) e^{-{1\over 2}( {\vec x} K {\vec x}^T+
{\vec y} K {\vec y}^T )} \! A_{ij} \partial_{x_i} \! \big( x_j \delta({\vec x}-{\vec y})\big)  \ ,
\end{equation}
with $A={(-\!1)^{a} \over a } (S^{aT}\!-\!S^{a})$. This simplifies to
\begin{equation}
\langle {\bf Q}_{2a} \rangle_{1k} = 4i \rho^2 \int \dd {\vec x} ( {\vec v}_k^* K {\vec x}) \Big( {\vec v}_k K A {\vec x}- ({\vec v}_k K {\vec x}) \, {\vec x} KA {\vec x} \Big) e^{-{\vec x} K {\vec x}^T} \ .
\end{equation}
Upon integration, the second term in the parenthesis gives a vanishing contribution and we obtain
\begin{equation}
\langle {\bf Q}_{2a} \rangle_{1k} =-i \, {{\vec v}_k^*  A K {\vec v}_k \over {\vec v}_k^*  K {\vec v}_k }= %-i  \, {\vec v}_k^*  A {\vec v}_k =
{(-\!1)^{a} \over a } \, 2 \sin p_k a \ ,
\end{equation}
where we have used that ${\vec v}_k$ is also an eigenvector of $K$. Along the same lines, the odd charges lead to
\begin{equation}
\langle {\bf Q}_{2a+1} \rangle_{1k} = {4 \rho^2 \over 2a\!+\!1} \int \dd {\vec x} ( {\vec v}_k^* K {\vec x}) \Big( 2 {\vec v}_k K^{2a+2} {\vec x}+({\vec v}_k K {\vec x}) \, {\rm tr} K^{2a+1} \Big) e^{-{\vec x} K {\vec x}^T} \ .
\end{equation}
The second term in the parenthesis is the vacuum contribution to the vev of the odd charges. Integrating we get
\begin{equation}
\langle {\bf Q}_{2a+1} \rangle_{1k} =  {2 \over 2a\!+\!1} \big({\vec v}_k^*  K^{2a+1} {\vec v}_k + {1 \over 2} {\rm tr} K^{2a+1} \big)= {2 \over 2a\!+\!1} \big( \omega_k^{2a+1}+ {1 \over 2} \sum_{l=1}^{L-1} \omega_l^{2a+1} \big)\ .
\end{equation}

\section{Appendix D}

In this appendix we find a solution $g(\theta)$  of eqs.\eqref{s10}.
The method will be later generalized to the elliptic  model.
In what follows we shall use the variable
\beq
z = \frac{\theta}{2 \pi i} \ .
\label{z}
\eeq
We first investigate  the periodicity properties of $g(z)$.
Combining the unitarity \eqref{s19}  and crossing symmetry relations \eqref{s10} one gets
\begin{equation}
g(z+1) =   \tan^4(\pi z)  g(z)  \label{ap1} \ ,
\end{equation}
 that implies that $g(z)$ is not a periodic function in $z$ (or periodic in $\theta$ with period $2 \pi i$).
Some  $S$-matrices exhibit this  periodicity that is equivalent to the  double
sheet structure in the $s$-plane, where $s$ is the Mandelstam variable \cite{M00}.  In our case, the lack of periodicity
implies  that the $S(s)$ has an infinite number of sheets in the $s$ complex plane.

Let us define the function
\barray
h(z ) = \lim_{M \rightarrow \infty}  \prod_{n=-M}^M h_n(z),  \quad  h_n(z) =
\frac{  \Gamma( n + z ) \Gamma \left( n +  \frac{1}{2} - z \right)     }{   \Gamma( n - z )  \Gamma \left( n +  \frac{1}{2}  + z \right) } \ ,
 \label{ap2}
\earray
where $h_n(z)$ satisfies
\beq
h_n\left(  \frac{1}{2} - z \right)  =     \frac{ n - \frac{1}{2} + z}{ n -z}  h_n(z) \ .
\label{ap3}
\eeq
Using
\beq
\tan( \pi z) =  \lim_{M \rightarrow \infty} \prod_{n = - M}^M  \frac{ n -z}{ n - \frac{1}{2} + z}    \ ,
\label{ap4}
\eeq
we find  that

 \beq
h\left(  \frac{1}{2} - z \right)   =     \frac{h(z)}{ \tan( \pi z) }    \longrightarrow
h^2\left(  \frac{1}{2} - z \right)   =     \frac{h^2(z) }{ \tan^2( \pi z) }   \, .
\label{ap5}
\eeq
which gives a solution of the  crossing symmetry relation satisfied by $g(z)$.

\beq
g(z) = h^2(z)  = \lim_{M \rightarrow \infty}
 \prod_{n=- M}^M   \left(   \frac{  \Gamma( n + z ) \Gamma \left( n +  \frac{1}{2} - z \right)     }{   \Gamma( n - z )  \Gamma \left( n +  \frac{1}{2}  + z \right) }
\right)^2  \ ,
\label{ap6}
\eeq
and the equations \eqref{s16}-\eqref{s10}.
The product form of \eqref{ap6} is a regularization of the function $g(z)$. If we naively replace the limit by
an  infinite product then \eqref{ap6}  would  be invariant under the replacement $z \rightarrow z+1$,
in contradiction with  \eqref{ap1}.

Next,  we find an expression for $\log g(z)$  using the formula \cite{AS}
\beq
\log \Gamma(x) = \left( x - \frac{1}{2} \right) \log x - x + \frac{1}{2} \log( 2 x) + \int_0^\infty \frac{dt}{t}  e^{ - t z} \left( \frac{1}{2} - \frac{1}{t} + \frac{1}{e^t - 1}\right) \ ,
\quad {\rm Re} \; x > 0  \ .
\label{ap7}
\eeq
This equation can be applied  in \eqref{ap6}  to  the terms where $ n >  0$  but not for those where $n  < 0$.
However, the latter terms can be transformed into the former ones using the relations
\beq
\Gamma(z+1) = z \Gamma(z), \qquad \Gamma(z) \Gamma(1 - z) = \frac{\pi}{\sin (\pi z)} \, ,
\label{ap8}
\eeq
that allow us to write
\beq
\prod_{n=-M}^M h_n(z) =  (-1)^{M+1} \frac{  \Gamma( \frac{1}{2} - z ) \Gamma \left( M+1 +   z \right)     }{  \Gamma( \frac{1}{2} +  z ) \Gamma \left( M+1 - z \right)   }    \prod_{n=1}^M h_n^2(z)  \, .
\label{ap9}
\eeq
%Equation \eqref{ap7} can be used for every factor in \eqref{ap9}, provided $|{\rm Re} \; z| <  \frac{1}{2}$.
Performing the sum over $n$ and taking the limit $M \rightarrow \infty$ yields
\beq
\log g(z)  =   \int_0^\infty \frac{dt}{t}  \frac{ \sinh( t z)}{ \cosh^2(t/4)}  , \qquad |{\rm Re} \; z| <  \frac{1}{2} \ .
\label{ap10}
\eeq
An alternative expression is obtained
taking the derivative respect to $z$
\beq
\frac{d \log g(z)}{dz }   =  \int_0^\infty dt   \frac{ \cosh( t z)}{ \cosh^2(t/4)}   = \frac{ 8 \pi z}{\sinh (2 \pi z)} .
\label{ap11}
\eeq
and integrating back

\beq
\log g(z)  =
 \frac{ i \pi}{2} + 4 z \log \frac{ 1 - e^{ 2 \pi i z}}{ 1 + e^{ 2 \pi  i z}} + \frac{ 2 i}{\pi} \left( {\rm Li}_2( -  e^{ 2 \pi i z}) - {\rm Li}_2(  e^{ 2 \pi i z} )   \right)  \ ,
 \label{ap12}
 \eeq
where ${\rm Li}_2(x)$ is a particular case of the  polylogarithmic function defined by  the analytic extension of the series
\beq
{\rm Li}_s(z) = \sum_{n=1}^\infty \frac{z^n}{n^s} \ ,  \qquad |z| < 1 \ ,
\label{ap13}
\eeq
to the complex plane. For $z=1$  it becomes the Riemann zeta function,  that is ${\rm Li}_s(1) = \zeta(s)$.

In the massive model, the function ${g}(z, \mu)$,  can be found  employing the previous  results.
Indeed, the crossing symmetry relation  satisfied by  ${g}(z, \mu)$ reads
  \beq
  {g}\left(  \frac{1}{2} - z, \mu  \right) =    \frac{1}{ c^2(z, \mu)}  {g}(z, \mu)  \ .
  \label{ap14}
 \eeq
 where
    \beq
 c(z, \mu) =  \sqrt{\mu_1}  \frac{ {\rm sn}(2 K z)}{{\rm cn}(2 K z)  {\rm dn}(2 K z)  }    \ ,
 \label{ap15}
 \eeq
with $K=K(\mu)$ defined in \eqref{K}. In the trigonometric case it becomes $c(z, 0) = \tan( \pi z)$.  The function \eqref{ap15}  satisfies
%$c(z, \mu)= - c \left( z \pm   \frac{ i m K' }{2 K} , \mu \right)$
$c(z, \mu)= - c \left( z \pm   \frac{ i m \tau }{2 \pi} , \mu \right)$, where $\tau={\pi K' \over K}$ and $K'=K(\mu_1)$.
It is  related to $c(z, 0)$ as
    \beq
 c(z, \mu) =  \lim_{M \rightarrow \infty}  \prod_{m=- M}^M  \tan \left[  \pi
 %\left( z + \frac{i m K'}{2 K}  \right) \right]   \ .
 \left( z + \frac{i m \tau}{2 \pi}  \right) \right]   \ .
 \label{ap15b}
 \eeq
 This yields a solution for   $g(z, \mu)$ in terms of  trigonometric $g$ function  \eqref{ap6}
 %Comparing these equations with those of the massless case we take
 %
 \beq
 {g}(z, \mu) = \lim_{M \rightarrow \infty}   \prod_{m= - M}^M  g
 %\left( z + i m \frac{K'}{2 K} \right)  \ ,
 \left( z +\frac{ i m  \tau}{2 \pi} \right)  \ ,
 \label{ap16}
 \eeq
that  using \eqref{ap6} becomes
 \begin{equation}
 {g}(z, \mu)   =    \lim_{M \rightarrow \infty}
 \prod_{n= - M}^M  \prod_{m= - M}^M   \left[
% \frac{ \Gamma \left( n + i m \frac{ K'}{2K} + z \right) \Gamma \left( n + \frac{1}{2} + i m \frac{ K'}{2K}  - z \right) } {
% \Gamma  \left( n + i m \frac{ K'}{2K} -  z \right)  \Gamma \left( n + \frac{1}{2} + i m \frac{ K'}{2K}  + z \right)}  \right]^2   \ .
 \frac{ \Gamma \left( n + \frac{i m \tau}{2 \pi} + z \right) \Gamma \left( n + \frac{1}{2} +\frac{i m \tau}{2 \pi}  - z \right) } {
 \Gamma  \left( n + \frac{i m \tau}{2 \pi} -  z \right)  \Gamma \left( n + \frac{1}{2} + \frac{i m \tau}{2 \pi} + z \right)}  \right]^2   \ .
 \label{ap17}
  \end{equation}
Similarly, using \eqref{ap10} we get
\barray
\log  {g}(z, \mu) &  =  &  \lim_{M \rightarrow \infty}   \sum_{m = - M}^M  \log g\left( z + \frac{i m \tau}{2 \pi} \right)  \label{ap18} \\
& = &
% \frac{ 2 \pi K z}{ K'} + \sum_{n= 1 }^\infty   \frac{1}{ n} \frac{ \sinh( 4 \pi K z n/K')}{ \cosh^2( \pi K n/K')}  \, , \quad  | {\rm Re} \ , z | < \frac{1}{2} \, .
\frac{ 2 \pi^2}{ \tau} z + \sum_{n= 1 }^\infty   \frac{1}{ n} \frac{ \sinh( 4 n \pi^2 z /\tau)}{ \cosh^2( n \pi^2/\tau)}  \, , \quad  | {\rm Re} \, z | < \frac{1}{2} \, .
 \nonumber
  \earray

%  \section{Appendix E}

\begin{figure}[t!]
\vspace{.2cm}
\begin{center}
\includegraphics[width=0.4\textwidth]{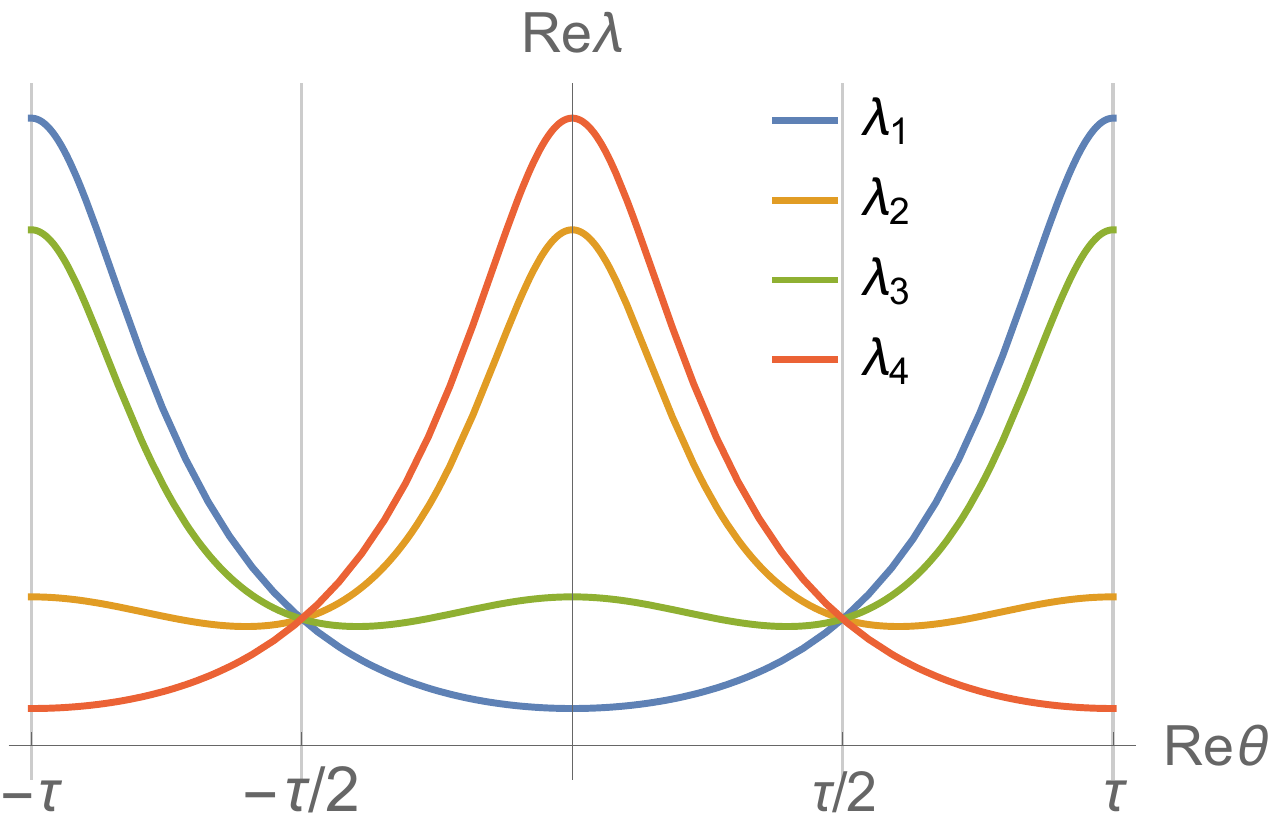} \qquad
\includegraphics[width=0.4\textwidth]{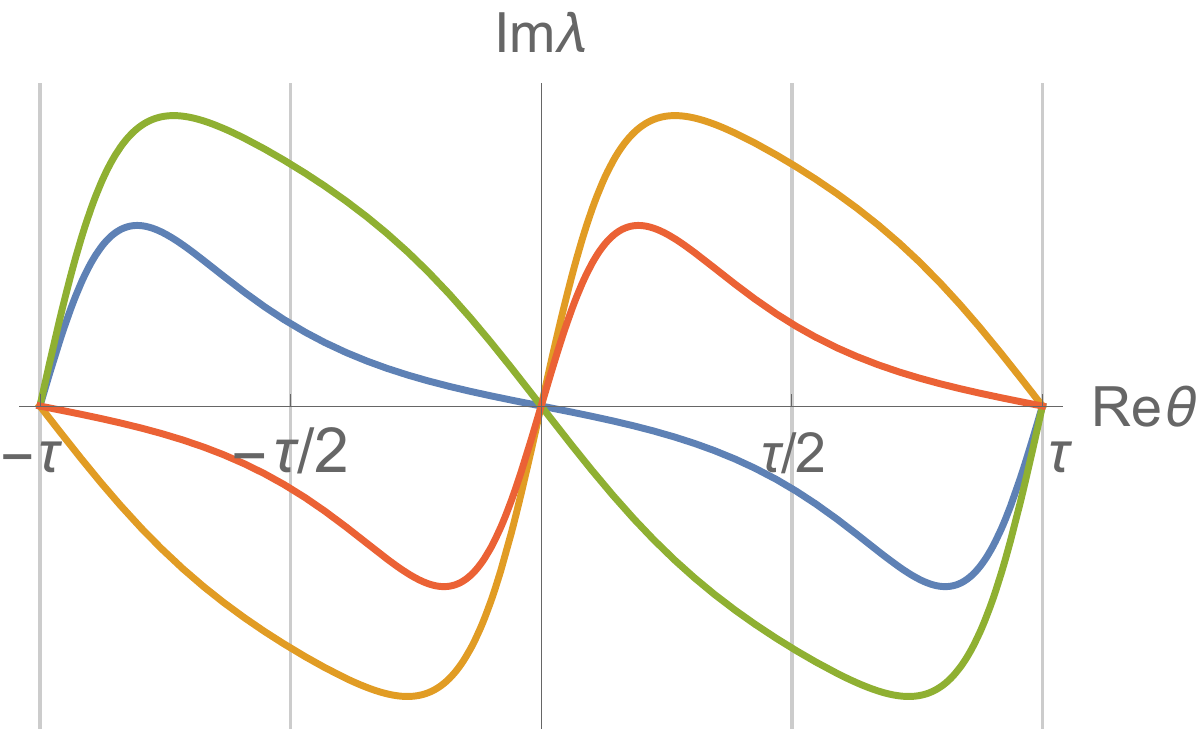}
\caption{Plot of the real and imaginary parts of the eigenvalues $\lambda_i$ for $\mu=0.3$ and ${\rm Re}\, z=0.05$.}
\label{eigen}
\end{center}
\end{figure}

Using this expression, we see that $g$ has the same periodicity along imaginary $z$ as $c$, namely $g(z, \mu)= - g \left( z \pm   \frac{ i m \tau }{2 \pi} , \mu \right)$.
Hence the prefactor in the elliptic S-matrix
\begin{equation}
S_{x_1x_2}^{y_1y_2}	= \frac{g}{2 \pi c }   \, e^{ -{c \over 2}   \bqty{ (x_1 - x_2)^2
+ (y_1 - y_2)^2 } -{1 \over 2 c}  \bqty{ (x_1 - y_1)^2 +  (x_2 - y_2)^2 }-{c \over 4}{\widetilde m}^2 \bqty{ (x_1^2 + x_2^2+y_1^2+y_2^2 ) }} \  ,
\label{Smmass}
\end{equation}
is invariant under that shift.
%satisfy the following property under a shift by $\tau={\pi K' \over K}$ in the rapidity
%\begin{equation}
%g(\theta+\tau,\mu)=-g(\theta,\mu) \ , \qquad  c(\theta+\tau,\mu)=-c(\theta,\mu) \ .
%\label{period}
%\end{equation}
%Hence the prefactor of the S-matrix has periodicity $\tau$.
The functions appearing on the exponent of \eqref{Smmass} have however twice that periodicity. In particular, the function $\widetilde m$ does not satisfy a simple half period relation like $c$ or $g$. This mismatch is better understood by diagonalizing the quadratic form in the exponent of the S-matrix
\begin{equation}
S_{x_1x_2}^{y_1y_2}	= \frac{g}{2 \pi c }   \, e^{ - \sum_{i=1}^4 \lambda_i e_i^2 } \ ,
\end{equation}
where
%. Denoting its eigenvalues by $-\lambda_i$ and their associated eigenvectors by $e_i$, we have
 \barray
 e_1={1 \over 2} (x_1+x_2+y_1+y_2)  \ , &&  \lambda_1= {c \over 4}\, {\widetilde m}^2 \ ,\\
 e_2={1 \over 2} (x_1+x_2-y_1-y_2)  \ , && \lambda_2= {1 \over c}+{c \over 4} \, {\widetilde m}^2 \ ,\\
e_3={1 \over 2} (x_1-x_2+y_1-y_2) \ , &&   \lambda_3= c+{c \over 4}  \, {\widetilde m}^2\ , \\
e_4={1 \over 2} (x_1-x_2-y_1+y_2)  \ ,\hspace{2cm}&&  \lambda_4= {1 \over c}+c +{c \over 4} \, {\widetilde m}^2 \ .
\earray
The real and imaginary part of $\lambda_i$ has been plotted in Fig.\ref{eigen} along a cycle $|{\rm Im} \, z| \leq {\tau \over 2 \pi}$.
In spite of the manifest ${\tau \over \pi}$ periodicity of the eigenvalues, an effective cyclicity with half the period emerges
by combining $z \to z\pm {i \tau \over 2 \pi}$ with the exchange $\lambda_1 \leftrightarrow \lambda_4$ and $\lambda_2 \leftrightarrow \lambda_3$.
This exchange just amounts to a change of sign in the continuous labels $x_2$ and $y_1$.

Consistency requires that the massive S-matrix has a well behaved limit when $|e_i| \to \infty$.
This implies that the eigenvalues $\lambda_i$ must always have a non-negative real part.
Fig.\ref{eigen} shows that this is indeed the case, and provides a non-trivial test since $c$ alone fails to fulfill this
requirement on half the ${\tau \over \pi}$ cycle.


\section*{References}

\begin{thebibliography}{99}

%%% BASIC REFERENCE ON EXACTLY SOLVABLE MODELS
\bibitem{B82}   R. J. Baxter,
{\em Exactly Solved Models in Statistical Mechanics}. Academic, London, 1982.

\bibitem{KB93} V. E. Korepin, N. M. Bogoliubov, A. G. Izergin,
{\em Quantum Inverse Scattering Method and Correlation Functions},
Cambridge Monographs on Mathematical Physics, Cambridge University Press, Cambridge, 1993.

\bibitem{G95} M. Gaudin,
{\em  Mod\`{e}les exactement r\'{e}solus}, Les \'{e}ditions de Physique,  France, 1995.

\bibitem{GR96} C. G\'omez, M. Ruiz-Altaba, G. Sierra,
{\em Quantum Groups in two-dimensional Physics},
Cambridge Monographs on Mathematical Physics, Cambridge University Press, London, 1996.


\bibitem{S04} B. Sutherland,
{\em Beautiful Models: 70 years of Exactly Solved Quantum Many-Body Problems},
World Scientific Publishing Co. Pte. Ltd., Singapore,  2004.

\bibitem{DP04} J. Dukelsky, S. Pittel, G. Sierra,
Exactly solvable Richardson-Gaudin models for many-body quantum systems,
Rev. Mod. Phys.{\bf 76}, 643  (2004).
%arXiv:nucl-th/0405011 [pdf, ps, other]

\bibitem{M10}  G. Mussardo,
{\em Statistical Field Theory:  An Introduction to Exactly Solved Models in Statistical Physics},
Oxford University Press, New York, 2010.


%%%%%%%%  OTHER METHODS

\bibitem{T95} A. M. Tsvelik,
{\em Quantum Field Theory in Condensed Matter Physics},
Cambridge University Press, Cambridge 1995.

\bibitem{GN98} A. O. Gogolin, A. A. Nersesyan, A. Tsvelik,
{\em Bosonization and Strongly Correlated Systems},
Cambridge University Press, Cambridge 1998.

\bibitem{G04} T.  Giamarchi,
{\em Quantum Physics in One Dimension},
Clarendon Press, Oxford, 2004.


\bibitem{W92} S. R. White,
Density matrix formulation for quantum renormalization groups,
 Phys. Rev. Lett. {\bf 69}, 2863 (1992).

\bibitem{S05}   U. Schollwock,
The density-matrix renormalization group,
Rev. Mod. Phys. {\bf 77}, 259 (2005).

\bibitem{O19} R. Or\'us, ``Tensor networks for complex quantum systems'', Nature Reviews Physics 1, 538 (2019).


%%%%%%%  STRING THEORY


\bibitem{MZ03} J. A. Minahan and K. Zarembo,
``The Bethe-ansatz for N = 4 super Yang-Mills'',
 JHEP 0303, 013 (2003).
 % [arXiv:hep-th/0212208].

\bibitem{BS03}  N. Beisert and M. Staudacher,
``The N = 4 SYM integrable super spin chain'',
Nucl. Phys. B {\bf 670}, 439 (2003)
%[arXiv:hep-th/0307042].


\bibitem{HL05} R.  Hernandez, E.  Lopez, A.  Perianez, G. Sierra,
``Finite size effects in ferromagnetic spin chains and quantum corrections to classical strings'',
JHEP 0506:011 (2005).
%arXiv:hep-th/0502188 [pdf, ps, other]


%%%%%%    Tensor Networks

\bibitem{MK12} V. Murg, V. E. Korepin, F. Verstraete,
The Algebraic Bethe Ansatz and Tensor Networks,
Phys. Rev. B 86, 045125 (2012).
%arXiv:1201.5627 [pdf, ps, other]


\bibitem{CM15} Y. Q. Chong, V. Murg, V. Korepin, F. Verstraete,
The nested Algebraic Bethe Ansatz for the supersymmetric t-J and Tensor Networks
 Phys. Rev. B {\bf 91}, 195132 (2015).
 %arXiv:1411.2839 [pdf, ps, other]

 %% elliliptic models

\bibitem{B72} R. J. Baxter,
``Partition function of the Eight-Vertex lattice model'',
Ann. Phys. {\bf 70}, 193 (1972).



\bibitem{Z79} A. B. Zamolodchikov,
$Z_4$ - Symmetric Factorized S-Matrix
in Two Space-Time Dimensions,  Commun. Math. Phys. {\bf 69}, 165  (1979).


\bibitem{M00} G. Mussardo, S. Penati,
``A Quantum Field Theory with Infinite Resonance States'',
Nucl.Phys. B{\bf 567},  454 (2000).
%   arXiv:hep-th/9907039.


%%%%%%%

\bibitem{CS19} M.  Campos, G. Sierra, E. L\'opez,
Tensor renormalization group in bosonic field theory,
 Phys. Rev. B {\bf 100}, 195106 (2019).

%%%%%


 \bibitem{AS} M. Abramowitz and I. Stegun,
 {\em Handbook of mathematical functions}, Dover Publications, Inc. New York 1972.

\bibitem{GP} A. Galindo and P. Pascual,
{\em Quantum Mechanics I (Theoretical and Mathematical Physics)},
 Texts and Monographs in Physics,
Springer-Verlag Berlin Heidelberg, 1990.



%%% Universal R

\bibitem{F96} L. D. Faddeev,
 ``How Algebraic Bethe Ansatz works for integrable model'',
 in {\em Sym\'etries Quantiques}, Proc. of
Les Houches School of Physics, Session LXIV, 1995, pp. 149-219 (North-Holland 1998).
hep-th/9605187.


\bibitem{D86} V.G. Drinfeld,
``Quantum Groups'', Proceedings of the 1986 International Congress of Mathematics at Berkeley ed. A.M. Gleason (1987) Am. Math. Soc., 1, p. 798.

\bibitem{CFT}  P. Di Francesco, P. Mathieu, D. S\'en\'echal, {\em Conformal Field Theory}, Springer, New York, 1997.

%%% other elliptic S matrices



\bibitem{ZZ79} A. B. Zamolodchikov and Al. Zamolodchikov,
Factorized $S$-matrices in two dimensions as the exact solutions
of certain relativistic quantum field theory models,
Ann. of Phys. {\bf 120}, 253 (1979).


\bibitem{W71} K. G. Wilson, ``Renormalization group and strong interactions'', Phys. Rev. D{\bf 3}, 1818 (1971).


\bibitem{BS12a} V. V. Bazhanov, S.  M. Sergeev,
``A master solution of the quantum Yang-Baxter equation and classical discrete integrable equations'',
 Adv. Theor. Math. Phys. {\bf 16}, 65 (2012).
%arXiv:1006.0651 [math-ph]


\bibitem{BS12b}   V. V. Bazhanov and S. M. Sergeev,
 Elliptic gamma-function and multi-spin solutions of the Yang-Baxter equation,
 Nucl.Phys. B {\bf 856}, 475 (2012).
 % arXiv:1106.5874 [math-ph].


\bibitem{G01} D.  Gottesman, A.  Kitaev and J. Preskill,
``Encoding a qubit in an oscillator'',
Phys. Rev. A {\bf 64}, 012310, (2001).

\end{thebibliography}
\end{document}